\RequirePackage{fix-cm}
\documentclass[smallextended]{svjour3} 
\smartqed  
\usepackage{graphicx}
\usepackage{cite}
\usepackage{amsmath,amssymb,amsfonts}
\usepackage{algorithmic}
\usepackage{graphicx}
\usepackage{textcomp}
\usepackage{xcolor}
\usepackage[hidelinks,bookmarks=true]{hyperref}
\usepackage[numbered]{bookmark}
\usepackage[small,it]{caption}
\usepackage{tcolorbox}
\usepackage{tabularx} 
\usepackage[switch]{lineno}
\usepackage{titlesec}
\usepackage{hyperref}
\usepackage{lineno}
\usepackage{dirtytalk}
\usepackage{footnote}
\usepackage{verbatim}
\usepackage{commath}
\usepackage{alphalph}
\usepackage{enumitem}
\usepackage[T1]{fontenc}
\usepackage{subcaption}
\usepackage{dirtytalk}
\usepackage[labelformat=parens,labelsep=quad, skip=3pt]{caption}
\usepackage{color,soul}
\usepackage{amsmath,amssymb,amsfonts}
\usepackage{algorithmic}
\usepackage{textcomp}
\usepackage{comment}
\usepackage{graphicx}
\usepackage{tabularx,booktabs}
\usepackage{array}
\usepackage{bbding}
\usepackage{pgf-pie}
\usepackage{pifont}
\usepackage{natbib}
\usepackage{wasysym}
\usepackage{amssymb}
\usepackage{amsmath} 
\usepackage{pgfplots}
\usepackage{sfmath}
\usepackage{array, booktabs, makecell}
\usepackage{color}
\usepackage{csquotes}
\usepackage{url}
\usepackage{comment}
\usepackage{tikz}
\usepackage{balance}
\usepackage{listings}
\usepackage{booktabs} 
\usepackage{soul} 
\usetikzlibrary{fit} 
\usetikzlibrary{positioning}
\usetikzlibrary{arrows}
\usetikzlibrary{shapes.multipart}
\usepackage{caption}
\usepackage{graphicx}
\usepackage{adjustbox}
\usepackage{float}
\usepackage{rotating}
\usepackage{tablefootnote}
\usepackage{nth}
\DeclareCaptionType{TextBox}
\newcolumntype{L}{>{\centering\arraybackslash}m{3cm}}
\usepackage[justification=centering]{caption}
\DeclareCaptionType{Quote}
\usepackage{color, colortbl}
\usepackage[utf8]{inputenc}
\usepackage{dirtytalk}
\usepackage{tcolorbox}
\usepackage{anyfontsize}
\usepackage{enumitem}
\usepackage{csquotes}

\usepackage{amsmath,amssymb,amsfonts}
\usepackage{algorithmic}
\usepackage{graphicx}
\usepackage{textcomp}
\usepackage{xcolor}
\usepackage[hidelinks,bookmarks=true]{hyperref}
\usepackage[numbered]{bookmark}
\usepackage{soul}
\usepackage{booktabs}
\usepackage{multirow}
\usepackage{float}
\usepackage{url}
\usepackage{tcolorbox}
\usepackage{tabularx} 
\usepackage[switch]{lineno}
\usepackage{titlesec}
\usepackage{balance}

\usepackage{xspace}

\newcommand{\RQone}{How effective is our supervised learning in predicting the type of refactoring?\xspace}
\newcommand{\RQtwo}{How do our model compare with keyword-based classification?\xspace}
\newcommand{\RQthree}{What are the frequent terms utilized by developers when documenting refactoring types?\xspace}
\newcommand{\RQfour}{How useful is our approach in analyzing the inconsistency types between source code and  documentation?\xspace}
\newcommand{\projectsNumber}{800\xspace}
\newcommand{\datasetSize}{5,004\xspace}
\newcommand{\typesNumber}{6\xspace}
\newcommand{\ie}{\textit{i.e.,}\xspace}
\newcommand{\eg}{\textit{e.g.,}\xspace}

\begin{document}

\title{\LARGE On the Documentation of Refactoring Types
}


\author{Eman Abdullah AlOmar \and Jiaqian Liu  \and Kenneth Addo  \and Mohamed Wiem Mkaouer \and 
        Christian Newman \and Ali Ouni \and Zhe Yu}
        
\authorrunning{AlOmar et al.}

\institute{%
  Eman Abdullah AlOmar 
  \at
  Rochester Institute of Technology \\
  \email{eman.alomar@mail.rit.edu}
    \and 
   Jiaqian Liu 
   \at
   University at Buffalo \\
  \email{jliu275@buffalo.edu} 
  \and
Kenneth Addo
\at 
University of Maryland \\
\email{kaddo1@umbc.edu}
  \and Mohamed Wiem Mkaouer \and Christian Newman \and Zhe Yu 
  \at 
  Rochester Institute of Technology \\
  \email{\{mwmvse,cdnvse,zxyvse\}@rit.edu} 
  \and
  Ali Ouni
  \at
  ETS Montreal, University of Quebec \\
  \email{ali.ouni@etsmtl.ca}
}

\date{Received: date / Accepted: date}

\maketitle

\begin{abstract}
Commit messages are the atomic level of software documentation. They provide a natural language description of the code change and its purpose. Messages are critical for software maintenance and program comprehension. Unlike documenting feature updates and bug fixes, little is known about how developers document their refactoring activities. Specifically, developers can perform multiple refactoring operations, including moving methods, extracting classes, renaming attributes, for various reasons, such as improving software quality, managing technical debt, and removing defects. Yet, there is no systematic study that analyzes the extent to which the documentation of refactoring accurately describes the refactoring operations performed at the source code level. Therefore, this paper challenges the ability of refactoring documentation, written in commit messages, to adequately predict the refactoring types, performed at the commit level. 
Our analysis relies on the text mining of commit messages to extract the corresponding features (\textit{i.e.}, keywords) that better represent each class (\textit{i.e.}, refactoring type). The extraction of text patterns, specific to each refactoring type (\textit{e.g.}, rename, extract, move, inline, etc.) allows the design of a model that verifies the consistency of these patterns with their corresponding refactoring. Such verification process can be achieved via automatically predicting, for a given commit, the method-level type of refactoring being applied, namely \textit{Extract Method}, \textit{Inline Method}, \textit{Move Method}, \textit{Pull-up Method}, \textit{Push-down Method}, and \textit{Rename Method}. We compared various classifiers, and a baseline keyword-based approach, in terms of their prediction performance, using a dataset of \datasetSize commits. Our main findings show that the complexity of refactoring type prediction varies from one type to another. \textit{Rename method} and \textit{Extract method} were found to be the best documented refactoring activities, while \textit{Pull-up Method}, and \textit{Push-down Method} were the hardest to be identified via textual descriptions. Such findings bring the attention of developers to the necessity of paying more attention to the documentation of these types. 
\end{abstract}


\section{Introduction}
\label{sec:Introduction}
Understanding maintenance activities is critical for practitioners to effectively support the evolution of their projects in terms of enhancing cost-effectiveness, managing technical debt, and better allocation of maintenance related resources. Therefore, a plethora of studies have been performed on automatic classification of repository artifacts (\textit{e.g.}, bug reports, issues, code changes) in general, and commit messages in particular for several purposes, including the approximation of maintenance activities \citep{gharbi2019classification,honel2020using}, identification of bug fixes \citep{zafar2019towards}, detection of security-relevant changes \citep{sabetta2018practical,alsolai2020systematic}. Recently, there have been a focus on analyzing commit messages in the context of refactoring.

Refactoring, being the art of improving software internal design without altering its external behavior \citep{alomar2021preserving}, is the \textit{de-facto} way to reduce technical debt \citep{avgeriou2016managing}. 
To help manage this technical debt, a lot of research focus has shifted to analyzing developers' refactoring practices through mining code changes and commit messages \citep{veerappa2013empirical,naiya2015relationship,ubayashi2018can,counsell2018developers,counsell2019relationship}. For instance, \citep{alomar2019can} developed a taxonomy of textual patterns, used by developers when documenting their refactoring activities, to understand how developers document these refactoring activities and many empirical studies have focused on mining commit messages to extract the reason behind developers' choice to refactor in terms of optimizing structural metrics, (\textit{e.g.}, coupling, complexity, etc.) \citep{pantiuchina2018improving,alomar2019impact}, and quality attributes (\textit{e.g.}, readability, etc.) \citep{fakhoury2019improving}. Commit messages were also used by \citep{rebai2020recommending} to recommend refactoring operations. 
While there is a heavy reliance on the valuable information contained in commit messages, little is known about the extent to which such information can properly describe the actual refactoring changes in the source code. Specifically, studies have shown that developers do often misuse refactoring related terminology, in their documentation \citep{zhangpreliminary18}. Because commit message analysis relies on the notion that refactorings are described in such a way that they can be distinguished from one another (\ie rename is described differently than move method), it is important to know whether this is generally true and in particular \textit{how} refactorings can be distinguished by the way they are described in commit messages.

\textcolor{black}{Recent studies have been heavily investigating how developers document refactoring to gain more insights on how refactoring is being practically applied. They parse commit messages to extract the intent behind the refactoring, then measure the impact of the refactoring on the source code quality, and verify the consistency between what was described in the message with the measurement in the source code. For instance, \citep{pantiuchina2018improving} found a misperception between the state-of-the-art structural metrics, widely used as indicators for refactoring, and what developers actually document as an improvement when they refactor their source code. Similarly, 
\citep{alomar2019impact} have found that not all metrics are equally capturing developers perception of software quality. \citep{fakhoury2019improving} have found that current readability frameworks are unable to capture what developers intended to be refactorings that improve the source code readability. Such misperception between the theory of detecting refactoring opportunities, through removing code smells and improving structural metrics, and practical intents driving developers to refactor, could explain the shortage of developers adoption of current refactoring tools \citep{murphy2012we,kim2014empirical}. \textcolor{black}{\citep{arnaoudova2016linguistic} investigated} the linguistic antipatterns that are in disjunction with the source code. \textcolor{black}{In another} important dimension that can be investigated, is the consistency between the documentation of the refactoring actions, and the refactoring types that were actually performed in the source code. Just like documenting features and bug fixes, recent studies have shown that developers intentionally describe refactoring activities in commit messages, \ie self-affirm the existence of refactoring activity \citep{alomar2019can,zhangpreliminary18}. Yet, little is known about the extent to which, the description of refactorings, in the commit message, matches the actual refactoring action that was committed.}  

\begin{figure}[]
 	\centering
 	\includegraphics[width=1.0\linewidth]{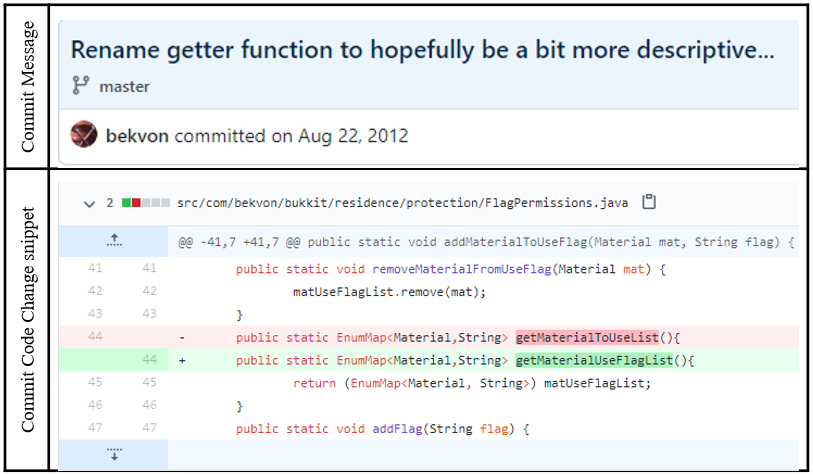}
 	\caption{\textcolor{black}{An example of a refactoring, and its corresponding documentation.}}
 	\label{fig:example}
\end{figure}

Therefore, we study the ways in which terminology used to describe refactorings in commit messages to  distinguish different refactorings from one another by studying the discriminative power of various machine learning techniques when provided this terminology. As an illustrative example, we refer to the simplified example extracted from the \texttt{bekvon/residence} project\footnote{ \url{https://github.com/bekvon/residence/commit/76c364ea47e5a28b2041a0bb3323cb48bab180c9} (last checked 2020/06/20)} reported in Figure \ref{fig:example}. The commit message states the purpose of refactoring as a rename of getter function for better readability. Based on the developers commit message, can we automatically deduce the existence of a refactoring whose type is \textit{Rename Method}. An intuitive solution for this problem is to detect the refactoring type in the source code and string-match it in the commit message to check whether it is mentioned as a form of verification. Such a solution assumes that developers refer to refactorings as they are known in the refactoring catalogue \citep{Fowler:1999:RID:311424,wake2004refactoring}. Previous studies found that developers misuse refactoring related terms \citep{soares2013comparing}, which hinders the accuracy of the string matching solution and presents a challenge for any solution that attempts to verify the consistency between refactoring and its corresponding documentation.

\textcolor{black}{The goal of our study is to investigate whether different words and phrases found in refactoring commit messages are unique to different types of refactorings (\textit{e.g.}, rename, move, extract, inline, etc.). In pursuit of this goal, we deploy machine learning techniques for the prediction of refactoring operation types based on commit messages. The results of this study can help us determine the types of words and phrases which best discriminate one type of refactoring from another; providing greater insight into the way refactoring is affirmed, which can be used to help automatically document refactorings in a more systematic way. Additionally, this work is critical in supporting refactoring documentation and in reducing the amount of effort needed by developers to appropriately describe what happened during a sequence of changes and help improve comprehension of those changes via commit messages. The work helps us understand how developers discriminate against different refactoring types through human language descriptions. Further, a recent industrial case study at Xerox reveals that developers rarely report specific refactoring operations as part of their documentation when submitting refactoring changes \citep{alomar2021icse}. \textcolor{black}{With} the lack of refactoring documentation guidelines, the reviewers are forced to ask for more details in order to recognize the need for refactoring. The authors designed a procedure for documenting any refactoring review requests, respecting three dimensions that they referred to as the three \textit{\textbf{I}}s, namely, \textit{\textbf{I}ntent}, \textit{\textbf{I}nstruction}, and \textit{\textbf{I}mpact}. Our study sheds light on the need to improve the quality of documenting refactoring types, which is considered one of the recommended dimensions to include in refactoring documentation.}


In this paper, we formulate the prediction of refactoring operation types as a multi-class classification problem. Our solution relies on textual mining of commit messages to extract the corresponding features (\textit{i.e.}, keywords) that better represent each class (\textit{i.e.}, refactoring type) in order to automatically predict, for a given commit, the type of refactoring being applied and documented. 

To build our model, we collected a dataset of commits that are known to contain the type of refactorings considered in this study. So, we use Refactoring Miner \citep{tsantalis2018accurate} to extract, from different open source projects, commits that are known to contain a refactoring operation. 
Using Refactoring Miner, we collected a dataset of \datasetSize instances, from \projectsNumber projects each instance represents a commit message, and a refactoring operation whose type is one of the \typesNumber method-level types considered in this study, namely \textit{Extract Method}, \textit{Inline Method}, \textit{Move Method}, \textit{Pull-up Method}, \textit{Push-down Method}, and \textit{Rename Method}. Then, we use the N-Gram technique \citep{manning1999foundations} to identify relevant features, for each of the classes, and which will be used to develop various classifiers, including Random Forest, Logistic Regression, and Gradient Boosted Machine.

Our key findings show that there is no uniform accuracy across all refactoring types, \ie some refactorings can achieve up to 90\% in terms of F-measure, while others achieve 35\% at best. This indicates that the documentations of some refactoring types, such as \textit{Rename Method} are likely to follow best documentation practices 
than others, while some types are harder to distinguish and tend to be more ambiguous 
, such as \textit{Move Method}, \textit{Pull-up Method}, and \textit{Push-down Method}. 

\textcolor{black}{This paper makes the following contributions:}

\begin{enumerate}

\item \textcolor{black}{
We identify the common keywords and phrases developers utilize when describing their refactoring activity in the commit messages. Since there is also a significant amount of ambiguity in the way words are used, our work can reduce this confusion and the keywords that we discuss in this work are a strong starting point for determining what phrases should be used to reduce ambiguity and improve the quality of refactoring documentation. To the best of our knowledge, this is the first work attempted to assess the quality of the documentation of refactoring types using text mining technique.}

\item \textcolor{black}{We formulate the refactoring type prediction as a multi-class classification problem based on commit messages mining, and we challenge various models.}

\item \textcolor{black}{We evaluate the performance of our prediction model by comparing it against a baseline keyword-based approach that relies on matching messages with known refactoring type  \citep{kim2014empirical,zhangpreliminary18,Ratzinger:2008:RRS:1370750.1370759,stroggylos2007refactoring,citeulike:2881658,soares2013comparing}}. 

\item \textcolor{black}{We discuss the inconsistency cases between the documentation of the refactoring actions, and the refactoring types that were actually performed in the source code.}

\item \textcolor{black}{We deploy our model as a lightweight web-service that is publicly available for software engineers and practitioners. We publicly provide our best model and the dataset that served as the \textit{ground-truth}, for replication and extension purposes \citep{SAR2019WEB}}\textcolor{black}{)}.

\end{enumerate}

The rest of this paper is structured as follows. We review existing studies related to refactoring documentation and commit classification in Section \ref{sec:RelatedWork}. 
Next, in Section \ref{sec:methodology}, we detail our classification methodology, including the data collection and preprocessing, and choice of the classification algorithms. Then, we evaluate our approach, in Section \ref{sec:results}, and report a comparative study between various classifiers, while identifying most influential features. In Section \ref{sec:Implication}, we report the implications of our study, and in Section \ref{sec:threats}, we discuss the threats to our work's validity. Finally, we conclude
and describe our future work in Section \ref{sec:conclusion}.

\section{Related Work}
\label{sec:RelatedWork}

\begin{table*}
  \centering
	 \caption{\textcolor{black}{Related Work in Commit Classification Using Machine Learning.}}
	 \label{Table:Related_Work_in_Commit_Classification}
\begin{sideways}
\begin{adjustbox}{width=\textheight,totalheight=\textwidth,keepaspectratio}
\begin{tabular}{lcclllll}\hline
\toprule
\bfseries Study & \bfseries Year  &  \bfseries Binary / Multi-class &  \bfseries Category & \bfseries Machine Learning  & \bfseries Training Size & \bfseries Result    \\
\midrule
 \multirow{2}{*}{\citep{article}} & \multirow{2}{*}{2006} &  \multirow{2}{*}{No/Yes}  &  Swanson's category & \multirow{2}{*}{Naive Bayes} & \multirow{2}{*}{400}   & \multirow{2}{*}{Accuracy: 70\%} \\  
& &  & Administrative & \\ \hline 
\multirow{3}{*}{\citep{5090025}} & \multirow{3}{*}{2009} &  \multirow{3}{*}{No/Yes} &  \multirow{2}{*}{Swanson's category} & J48 / Naive Bayes / SMO  & \multirow{3}{*}{2000}  & \multirow{2}{*}{F-measure: 51\%}  \\ 
& & &  \multirow{2}{*}{Feature Addition} &  KStar / IBk / JRip / ZeroR & & \multirow{2}{*}{Accuracy: 52\%} \\ 
& & & & Non-Functional \\ \hline
\multirow{2}{*}{\citep{Hindle:2011:ATN:1985441.1985466}} & \multirow{2}{*}{2011} &  \multirow{2}{*}{No/Yes}  & \multirow{2}{*}{Non-Functional} &  rule / decision trees / vector space & \multirow{2}{*}{Not mentioned} & Receiver Operating  \\
&  & & & SVM / CLR / HOMER / BR 
&  & Characteristic up to 80\% 
\\ \hline
\citep{Levin:2017:BAC:3127005.3127016} & 2017 & No/Yes  & Swanson's category &  J48 / GBM / RF & 1151  & Accuracy: 76\% \\
\hline 
\multirow{3}{*}{\citep{honel2019importance}} & \multirow{3}{*}{2019} & \multirow{3}{*}{No/Yes}  & \multirow{3}{*}{Swanson's category} & LssvmRadical / SVM /  GBM     & \multirow{3}{*}{1151}  & \multirow{3}{*}{Accuracy: up to 89\%} \\ 
& &  & & xgbTree / LDA / MDA / NN / avNNet  & & \\ 
& & &  & C5.0 / RF / Naive Bayes / LogitBoost & & \\\hline
\citep{gharbi2019classification} & 2019 & No/Yes & Swanson's category & DT / kNN / RF / MLP & 5000  & F-measure: 45.79\%\\ \hline
\multirow{2}{*}{\citep{Jane2020enhancing}} & \multirow{2}{*}{2020} & \multirow{2}{*}{Yes/Yes} & binary: CMR vs non-CMR & \multirow{2}{*}{NB / LR / SVM / kNN} & \multirow{2}{*}{1529}  & F-measure: 84\%  \\
& & & multi-class: 12 refactoring types &  & & F-measure: 71\% \\ \hline
\multirow{2}{*}{\citep{alomar2020toward}} & \multirow{2}{*}{2020} & \multirow{2}{*}{Yes/Yes} & binary: SAR vs non-SAR & RF / LR / GBM / DJ / BPM & 1823  & F-measure: 98\%  \\
& & & multi-class: Internal QA / External QA / code smell & SVM / LD-SVM / NN / AP & 1044  & F-measure: 93\% \\ \hline
\multirow{2}{*}{\citep{alomar2021we}} & \multirow{2}{*}{2020} & \multirow{2}{*}{No/Yes} & multi-class: Internal QA / EXternal QA / code smell & RF / LR / kNN / DT / SVC  & \multirow{2}{*}{1702} &  \multirow{2}{*}{F-measure: 87\%} \\
& & & Bug Fix / Functional & Mutlinomial Naive Bayes \\ \hline
\textcolor{black}{\citep{aniche2020effectiveness}} & \textcolor{black}{2020} & \textcolor{black}{Yes/No} & \textcolor{black}{multi-class: 20 refactoring types} & \textcolor{black}{LR / NB / SVM / DT / RF / NN} & \textcolor{black}{2 million} & \textcolor{black}{F-measure:  > 90\%}  \\ \hline
\textcolor{black}{\citep{marmolejos2021use}} & \textcolor{black}{2021} & \textcolor{black}{Yes/No} & \textcolor{black}{SAR vs non-SAR} & \textcolor{black}{BPM / AP / LR / GBM / NN} & \textcolor{black}{3000} &  \textcolor{black}{F-measure: 96\%} \\ \hline
\textcolor{black}{\multirow{2}{*}{\citep{alomar2021behind}}} & \multirow{2}{*}{\textcolor{black}{2021}} & \multirow{2}{*}{\textcolor{black}{No/Yes}} & \textcolor{black}{multi-class: Internal QA / External QA / code smell} & \textcolor{black}{RF / LR / kNN / DT / SVC } & \multirow{2}{*}{\textcolor{black}{1702}} &  \multirow{2}{*}{\textcolor{black}{F-measure: 87\%}} \\
& & & \textcolor{black}{Bug Fix / Functional} & \textcolor{black}{Mutlinomial Naive Bayes} \\ 

\bottomrule
\end{tabular}
\end{adjustbox}
\end{sideways}
\vspace{-.2cm}
\end{table*}

In this section, we report studies related to developer's perception of refactoring and its documentation, along with the current state-of-the-art studies related to commit messages classification. 

\subsection{Refactoring Documentation}

A number of studies have focused recently on the identification and detection of refactoring activities during the software life-cycle. One of the common approaches to identify refactoring activities is to analyze the commit messages in versioned repositories. \citep{stroggylos2007refactoring} searched words stemming from the verb \textit{\say{refactor}} such as \say{refactoring} or \say{refactored} to identify refactoring-related commits.  \citep{Ratzinger:2008:RRS:1370750.1370759,citeulike:2881658} also used a similar keyword-based approach to detect refactoring activity between a pair of program versions to identify whether a transformation contains refactoring. The authors identified refactorings based on a set of keywords detected in commit messages, and focusing on the following 13 terms in their search approach: \textit{refactor, restruct, clean, not used, unused, reformat, import, remove, replace, split, reorg, rename, and move}. 

Later, \citep{murphy2012we} replicated Ratzinger's experiment in two open source systems using Ratzinger's 13 keywords. They conclude that commit messages in version histories are unreliable indicators of refactoring activities. This is due to the fact that developers do not consistently document refactoring activities in the commit messages. In another study, \citep{soares2013comparing} compared and evaluated three approaches, namely,  manual analysis, commit message (Ratzinger et al.'s approach \citep{Ratzinger:2008:RRS:1370750.1370759,citeulike:2881658}), and dynamic analysis (SafeRefactor approach \citep{Soares2009safetytool}) to analyze refactorings in open source repositories, in terms of behavioral preservation. The authors found, in their experiment, that manual analysis achieves the best results in this comparative study and is considered as the most reliable approach in detecting behavior-preserving transformations. 

In another study, \citep{kim2014empirical} surveyed 328 professional software engineers at Microsoft to investigate when and how they do refactoring. They first identified refactoring branches and then asked developers about the keywords that are usually used to mark refactoring events in commit messages. When surveyed, the developers mentioned several keywords to mark refactoring activities. \citep{kim2014empirical} matched the top ten refactoring-related keywords identified from the survey against the commit messages to identify refactoring commits from version histories. Using this approach, they found 94.29\% of commits do not have any of the keywords, and only 5.76\% of commits included refactoring-related keywords.

Prior works \citep{zhangpreliminary18,alomar2019can} have explored how developers document their refactoring activities in commit messages; this activity is called Self-Admitted Refactoring or Self-Affirmed Refactoring (SAR). In particular, SAR indicates developers' explicit documentation of refactoring operations intentionally introduced during a code change. The existence of such patterns unlocks more studies that question the developer's perception of quality attributes (\textit{e.g.,} coupling, complexity), typically used in recommending refactoring. For instance,  \citep{alomar2019impact} identified which quality models are more in-line with the developer's vision of quality optimization when they explicitly mention in the commit messages that they refactor to improve these quality attributes. This study shows that, although there is a variety of structural metrics can represent internal quality attributes, not all of them can measure what developers consider to be an improvement in their source code. \textcolor{black}{Furthermore, \citep{alomar2021behind} explored the relationship between developers' experience and refactoring. Their main findings show that refactoring contributors that frequently refactor the code tend to document less than developers that occasionally perform refactoring.}

\subsection{Commit Classification}
 \citep{5090025} proposed an automated technique to classify commits into maintenance categories using seven machine learning techniques. To define their classification schema, they extended Swanson's categorization \citep{Swanson:1976:DM:800253.807723} with two additional changes: Feature Addition, and Non-Functional. They observed that no single classifier is the best. \textcolor{black}{\citep{Hindle:2011:ATN:1985441.1985466} conducted} another experiment that classifies history logs in which their classification
of commits involves the non-functional requirements (NFRs) a commit addresses. Since the commit may possibly be assigned to multiple NFRs, they used three different learners for this purpose along with using several single-class machine learners.  \citep{article} had a similar idea to \citep{5090025} and extended the Swanson categorization
hierarchically. They, however, selected one classifier (\textit{i.e.,} Naive Bayes) for their classification of code transactions. Moreover, maintenance requests have been classified using two different machine learning techniques (\textit{i.e.,} Naive Bayesian and Decision Tree) in \citep{5561540}.  \citep{McMillan:2011:CSA:2117694.2119646} explored three popular learners to categorize software application for maintenance. Their results show that SVM is the best performing machine learner for categorization over the others. 

\citep{Levin:2017:BAC:3127005.3127016} automatically classified commits into three main maintenance activities using three classification models namely, J48, Gradient Boosting Machine (GBM), and Random Forest (RF). They found that the RF model outperforms the two other models (accuracy: 76\% versus 70\% and 72\%). 
 Recently, a replicated study \citep{honel2019importance} of \citep{Levin:2017:BAC:3127005.3127016} introduced code density of a commit to study the purpose of a change. Using code-density based classification, they achieved up to 89\% accuracy for cross project commit classification using LogitBoost classifier. In another study, \citep{gharbi2019classification} proposed a multi-label active learning-based approach to classify commit messages into maintenance categories. Their experimental results showed that the proposed approach achieved an F-measure of 45.79\%.  

\citep{Jane2020enhancing} developed a model to first detect refactoring commit messages from non-refactoring commits, and then differentiated between 12 refactoring types. Their findings showed that Naive Bayes and SVM achieved the best performance with an F-measure of 84\% and 0.71\% for binary and multiclass classification problems, respectively. \textcolor{black}{Another experiment that predicts refactoring was conducted using quality metrics. \citep{aniche2020effectiveness} used a machine learning approach that involves predicting refactoring using code, process, and ownership metrics. The resulting
models predict 20 different refactorings at class, method, and variable-levels with an accuracy often higher than 90\%}. More recently, \citep{alomar2020toward} proposed an approach to classify self-affirmed
refactoring in commit messages. Their results show that their approach is able to accurately classify SAR commits with accuracy of 98\% and 93\% for two-class and multiclass classification methods, respectively, outperforming the two state-of-the-art approaches, \ie the keyword-based and the random classifier. In a follow-up work, \citep{alomar2021we} performed a multi-class classification to categorize these commits into three categories, namely, Internal Quality Attribute, External Quality Attribute, and Code Smell Resolution, along with the traditional Bug Fix and Functional categories. This classification challenges the original definition of refactoring, being exclusive to improving software design and fixing code smells.  \textcolor{black}{\citep{marmolejos2021use} proposed a framework to identify refactoring documentation by using different techniques, such as
feature hashing and feature selection (Chi-squared and Fisher
score), and five machine learning algorithms.  As per their results, the combination of Chi-Squared with Bayes point machine and Fisher score with Bayes point
machine could be the most efficient when it comes to automatically identifying refactoring documentation, with an F-measure of 96\%.}  We summarize these state-of-the-art studies in Table~\ref{Table:Related_Work_in_Commit_Classification}. 


Our work is in the intersection of the above-mentioned studies, as we \textcolor{black}{leverage} commit classifications techniques to automatically classify refactoring documentation. While prior studies searched for the existence of refactoring documentation, we further challenge it by checking whether the granularity of documentation can reach up to the level of distinguishing the types of executed refactorings. \textcolor{black}{The refactoring types that we want to identify are the following:}

\begin{itemize}
\item \textcolor{black}{\textit{Extract Method.} creating a new method by extracting a selection of code from inside the body of an existing method.}
\item \textcolor{black}{\textit{Inline Method.} replacing calls and usages of a method with its body, and potentially removing its declaration.}
\item \textcolor{black}{\textit{Move Method.} changing the declaration of a method, from one class to another one.}
\item \textcolor{black}{\textit{Pull-up Method.} moving up a method in the inheritance chain from a child class to a parent class.}
\item \textcolor{black}{\textit{Push-down Method.} moving down a method in the inheritance chain from a parent class to a child class.}
\item \textcolor{black}{\textit{Rename Method.} changing the name of a method identifier to a different one.}

\end{itemize}

We chose types that are applied to the same level, \ie  for the sake of consistency. Our approach can also be applied to class-level or package-level refactorings. In the next section, we detail the design of our proposed approach.

 \section{Study Design}
\label{sec:methodology}

The aim of our work is to reveal the extent to which a clear documentation of refactorings can help in correctly classifying them. 
The manual search for such correlation between refactoring types and their corresponding proper description can be time-consuming and error prone.
We refer to solutions that can properly discriminate, and resolve, textual ambiguity; imitating the human decision making \citep{murphy2012machine} versus other, simpler techniques such as string-matching \citep{Ratzinger:2008:RRS:1370750.1370759,stroggylos2007refactoring,soares2013comparing,ratzinger2005improving} which can be used, to some extent, to solve the same problem. We opt for the supervised learning where predictors (\ie independent variables) are developed to decide about the dependent variable's value, which, in our case, refers to the commit message classification. Thus, our dependent variable is represented by the refactoring types to be predicted. The independent variables will be extracted from the keywords used by developers to describe each type of refactoring in their commit messages. Therefore, we need to first setup a dataset that can characterize each class adequately. Since our aim is to investigate which types of refactoring are more adequately documented than others, we formulate this problem as a multiclass classification problem. Hence, when we build our dataset, we choose commits such that each contains one type of refactoring being performed. 
Then, we provide, for each class (\ie refactoring type) a set of commit messages that are meant to document it. 

\begin{figure*}[htbp]
\centering 
\includegraphics[width=1\textwidth]{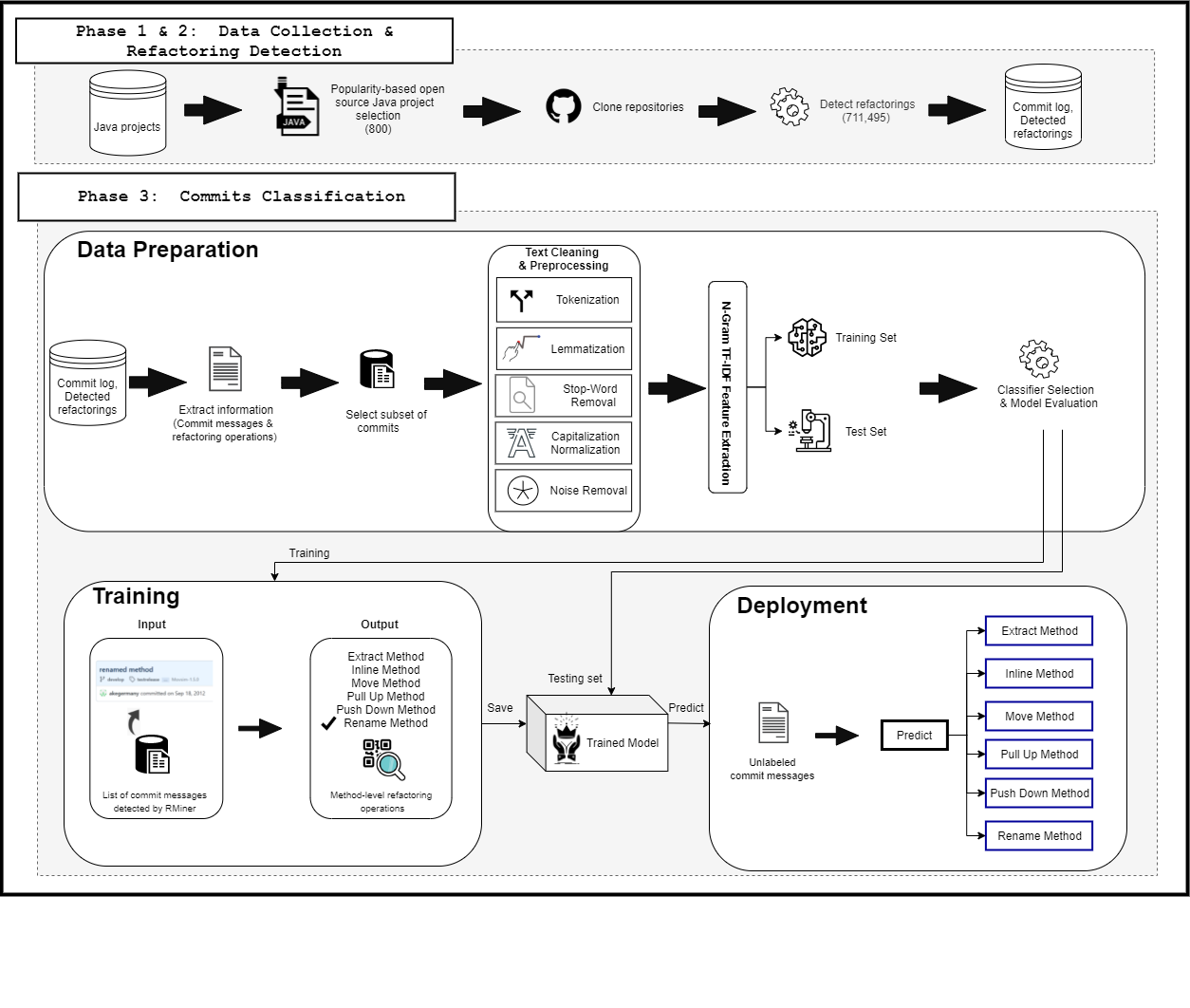}
\caption{Overall Prediction Framework.}
\label{fig:approach_overview}
\end{figure*}

In the following, we elaborate on the technical details of our adopted classification technique, starting from the data collection, through its preparation and finally the models training and validation. The overview of our approach is depicted in Figure~\ref{fig:approach_overview}.

\subsection{Overall Classification Framework}

In a nutshell, the goal of our work is to automatically identify then classify commit messages containing refactoring documentation. Our approach takes as input, a commit message, and  classifies it into one of six common method-level refactoring operations: \textit{Extract Method}, \textit{Inline Method}, \textit{Move Method}, \textit{Pull-up Method}, \textit{Push-down Method}, and \textit{Rename Method}. The overall framework of our approach is depicted in Figure~\ref{fig:approach_overview}. We formulate a two-phased approach that consists of a model building phase and a prediction phase. In the model building phase, our goal is to build a model from a corpus of real-world documented refactoring operations (\textit{i.e.,} commit messages). In the prediction phase, the built model will be used to predict the type of a given refactoring-related commit messages.

Our framework takes commit messages along with their ground truth categories obtained by manual inspection as input for the training procedure extracted from different projects. Based on this input, the commit messages are preprocessed, allowing for informative featurization. Thereafter, for each commit message, we extract features (\textit{i.e.,} words) to create a structured feature space. Then, we use the extracted features to build the training set. In total, \textcolor{black}{we experimented with 9 commonly} used classifiers to evaluate our prediction model, namely, Gradient Boosted Machine (GMB) \citep{friedman2001greedy}, Support Vector Machine (SVM) \citep{wu2008top}, Locally Deep SVM (LD-SVM) \citep{jose2013local}, Averaged Perceptron Method (APM) \citep{collins2002discriminative}, Bayes Point Machine (BPM) \citep{herbrich2001bayes}, Logistic Regression (LR) \citep{andrew2007scalable}, Random Forest (RF) \citep{prinzie2008random}, Decision Jungle (DJ) \citep{shotton2013decision}, and Neural Network (NN) \citep{hansen1990neural}. 
We selected these classifiers as they are commonly used in previous commit classification studies as well as several software engineering classification/prediction problems \citep{Hindle:2011:ATN:1985441.1985466,Levin:2017:BAC:3127005.3127016,article,5090025,5561540,levin2019towards,honel2019importance}, 
 as outlined in Table \ref{Table:Related_Work_in_Commit_Classification}. After training all models, we use a testing set to challenge the performance. Since the model has already learned the vocabulary of N-Gram (discussed in Section~\ref{sec:n-gram}) and their weights from the training dataset, we extract features from the test data based on that vocabulary and weights, and input them to the model. Finally, the classifier will output the predicted label for each tested commit message. 

\subsection{Commit Classification}
Our solution design has six main phases: (1) data collection and refactoring detection, (2) data labeling, (3) text cleaning and preprocessing, (4) feature extraction using N-Gram, (5) model training and building, and (6) model evaluation. Since a commit message is written in plain text, we follow the approach provided by \citep{kowsari2019text,alomar2020toward} that discussed a recent trend in text classification techniques and algorithms.  

\subsubsection{Data Collection \& Refactoring Detection}
To perform this study, we randomly selected 800 projects, which were curated open-source Java projects hosted on GitHub as described in  Table~\ref{Table:DATA_Overview}. These curated projects were selected from an available dataset by \citep{munaiah2017curating}, while verifying that they were Java-based; the only language supported by Refactoring Miner.
The authors of this dataset classified \say{well-engineered software projects} based on the projects’ use of software engineering practices such as documentation, testing, and project management. Additionally, these projects are non-forked (\ie not cloned from other projects), as forked projects may impact our conclusions by introducing duplicate code and documents. 
Also, 74.6\% of the projects had their most recent commit within the last four years. The 800 selected projects analyzed in this study have a total of 748,001 commits, and a total of 711,495 refactoring operations from 111,884 refactoring commits. 
Additionally, these projects contain 732 commits and involve 19 developers on average  \textcolor{black}{(corresponding to the median of 346.5 commits and 7 developers)}. An overview of the projects is provided in Table~\ref{Table:DATA_Overview}. 

To extract the entire refactoring history of each project, we use Refactoring Miner because it achieved the highest accuracy in detecting refactorings compared to the state-of-the-art available tools, with a precision of 98\% and recall of 87\% \citep{tsantalis2018accurate,silva2016we} along with being suitable for our study that requires a high degree of automation in data mining.

\begin{table}[h]
\begin{center}
\caption{\textcolor{black}{Projects Overview.}}
\label{Table:DATA_Overview}
\begin{tabular}{lr}\hline
\toprule
\bfseries Item & \bfseries Count \\
\midrule
Total of projects & 800 \\
Total commits & 748,001 \\
Refactoring commits & 111,884 \\
Refactoring operations & 711,495 \\
\midrule 
\multicolumn{2}{c}{\textbf{\textit{Considered Projects - Refactored Code Elements}}}\\
\bfseries Code Element & \bfseries \# of Refactorings  \\
\midrule
Method & 302,929 \\
Class & 228,974  \\
Attribute & 80,509 \\
Parameter & 42,992 \\
Variable & 28,765 \\
Package & 2380 \\
Interface & 1742 \\
\bottomrule
\end{tabular}
\end{center}
\end{table}

\subsubsection{Data Labeling}

Our goal is to provide the classifier with sufficient commits that represent the refactoring operations considered in this study. Since the number of candidate commits to classify is large, we cannot manually process them all, and so we need to randomly sample a subset while making sure it equitably represents the featured classes, \textit{i.e.,} refactoring types. Since an imbalanced training dataset or class starvation (\textit{i.e.,} not having adequate instances of a certain class) could worsen the performance of the model \citep{Levin:2017:BAC:3127005.3127016,levin2019towards}, we make sure that the classes for  multiclass classification problem are equally distributed when preparing the data for the training (\textit{cf.,} Table~\ref{Table:Instances per class (train, test)}). The classification process has been performed by the authors of the paper. To approximate the needed number of commits to add, we reviewed the thresholds used in the studies related to commit classification (see Table~\ref{Table:Related_Work_in_Commit_Classification}). The highest number of commits used in comparable studies was 5,000 commits \citep{gharbi2019classification}. Thus, we select a sample of 5,004 commits from 800 projects for each classification model. Below we detail the manual analysis of the data we use for our classification.

To prepare the dataset for the multiclass classification, we first run Refactoring Miner \citep{tsantalis2018accurate} on the \projectsNumber open-source projects we presented in Table \ref{Table:DATA_Overview}, in order to identify all commits containing refactorings. Then, we filter them to only keep commits with at most one refactoring type. Then, we cluster them by the types of refactorings we selected for this study. For each cluster, we start the random sampling of potential commits to include for our training set. For each randomly \textcolor{black}{selected} commit, we manually read through its message to verify whether it contains any textual description of the refactoring. Any commit with no such textual description is discarded. \textcolor{black}{In our work, we discard the commits that do not contain any textual description of refactoring to narrow down the commit messages eliminating the ones that are less likely to be classified as one of the refactoring types. It is important to note that we removed these commit messages because (1) these commit messages do not contain enough information and do not describe the code change , and (2) we want to train the classifier on well-documented commit messages, and label commits that contain enough information about refactorings so that we can assess the quality of refactoring documentation.}
An example of commits that we retain in our dataset is illustrated in Figure \ref{fig:example}. An example of commits that we discard documents a pull request, \eg "\textit{Merge pull request \#6 from marcel-blonk/develop make map type handle interfaces correctly}"\footnote{Commit extracted from sage-bionetworks/schema-to-pojo.}. Commits whose messages do not contain any kind of refactoring documentation would represent a noise in our dataset. Such commits would have been kept if the problem was formulated to binary identify refactoring documentation, but this is out of the scope of our work. This process resulted in selecting 5,004 stratified samples, divided equally for each stratum.

\begin{figure}[]
 	\centering
 	\includegraphics[width=1.0\linewidth]{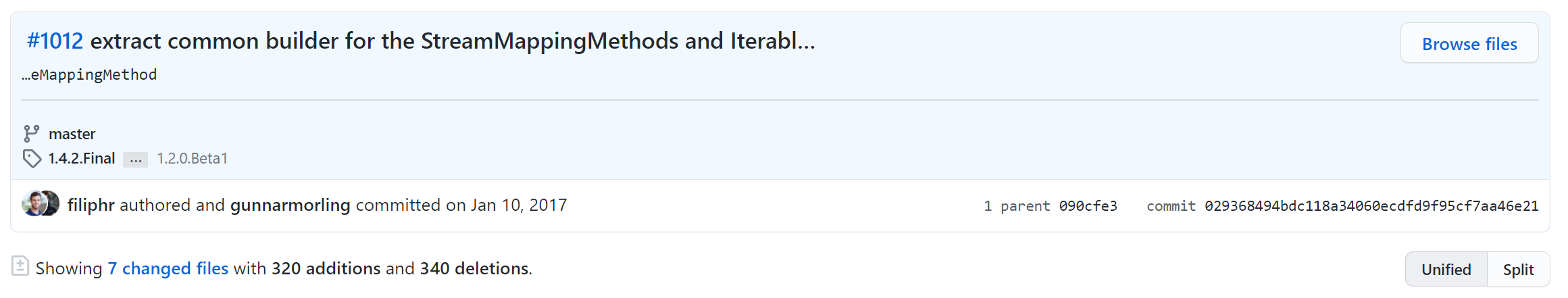}
 	\caption{\textcolor{black}{An example of multiple refactorings, and its corresponding documentation.}}
 	\label{fig:example_multilabel_case}
\end{figure}

\textcolor{black}{It is worth noting that upon performing the manual inspection of a subset of commit messages, we noticed that developers mostly document refactoring when they perform one or very few refactoring operations. However, if developers performed multiple refactoring operations, they are unlikely to detail refactoring activity in the commit messages. Figure \ref{fig:example_multilabel_case} depicts an example of a commit message in which a developer stated that they performed only Extract refactoring operations. Yet, when running the Refactoring Miner tool, it shows that there are 36 refactoring operations performed in this commit message, namely, \textit{Extract Method}, \textit{Extract Superclass}, \textit{Pull up Attribute}, \textit{Pull up Method}, and \textit{Rename Method}.}

\begin{table}[h]
\begin{center}
\caption{Number of Refactoring Instances per Class.}
\label{Table:Instances per class (train, test)}
\begin{adjustbox}{width=1.0\columnwidth,center}
\begin{tabular}{lcccccc}\hline
\toprule
\bfseries Dataset & \bfseries Extract & \bfseries Inline & \bfseries Move  & \bfseries Pull Up & \bfseries Push Down & \bfseries Rename  \\
\midrule
5,004 instances & 834 & 834 & 834 & 834 & 834 & 834  \\
\bottomrule
\end{tabular}
\end{adjustbox}
\end{center}
\end{table}

\subsubsection{Text Cleaning \& Preprocessing}
After the data preparation phase, we applied a similar methodology explained in \citep{kowsari2019text,kochhar2014automatic} for text pre-processing. In order for the commit messages to be classified into correct categories, they need to be preprocessed and cleaned; put into a format that the classification algorithms will process. This way, the noise will be removed, allowing for informative featurization. To extract features (\textit{i.e.,} words), we preprocess the text as follows: 

\begin{itemize}
    \item \textbf{Tokenization:} The goal of tokenization is to investigate the words in a sentence. The tokenization process breaks a stream of text into words, phrases, symbols, or other meaningful elements called tokens \citep{kowsari2019text}. In our work, we tokenize each commit by splitting the text into its constituent set of words. We also split tokens on special characters (\textit{e.g.,} the string \say{package-level} would be separated into two tokens, \say{\textit{package}} and \say{\textit{level}}).
    \item \textbf{Lemmatization:} The lemmatization process either replaces the suffix of a word with a different one or removes the suffix of a word to get the basic word form (lemma). We opted to use lemmatization over stemming, as the lemma of a word is a valid English word \citep{lane2019natural}. In our work, the lemmatization process involves sentence separation, part-of-speech identification, and generating dictionary form. We split the commit messages into sentences, since input text could constitute a long chunk of text. The part-of-speech identification helps in filtering words used as features that aid in key-phrase extraction. Lastly, since the word could have multiple dictionary forms, only the most probable form is generated.  
    \item \textbf{Stop-Word Removal:} Stop words (\textit{i.e.,} words and common English words such as \say{is}, \say{are}, \say{if}, etc) are removed since they do not play any role as features for the classifier \citep{saif2014stopwords}.  
    \item \textbf{Capitalization Normalization:} Since text could have a diversity of capitalization to form a sentence and this could be problematic when classifying large commits, all the words in the commit messages are converted to lower case and all verb contractions are expanded.
    \item \textbf{Noise Removal:} Special characters and numbers are removed since they can deteriorate the classification. More specifically, we remove all numeric characters, unique and duplicate special characters, email addresses and URLs. 
\end{itemize}

\subsubsection{Feature Extraction Using N-Gram} 
\label{sec:n-gram}
After cleaning and preprocessing the text, we apply feature extraction to extract only the most useful information from text strings to differentiate classes in both classification problems. In particular, we selected the N-Gram technique for feature extraction. The N-Gram technique is a set of \textit{n-word} that occurs in a text set and could be used as a feature to represent that text \citep{kowsari2019text}. In general, N-Gram term has more semantic than an isolated word. Some of the keywords (\textit{e.g.,} \say{\textit{extract}}) do not provide much information when used on its own. However, when collecting N-Gram from commit message (\textit{e.g.,} \textit{Refactor createOrUpdate method in MongoChannelStore to extract methods and make code more readable}), the keyword \say{extract} clearly indicates that this refactoring commit belongs to \textit{Extract Method} refactoring. 
In our classification, we use bigrams since it is very common to enhance the performance of text classification \citep{tan2002use}, and we select Fisher Score filter-based feature selection \citep{duda2012pattern,gu2012generalized} to \textit{featurize} text and manage the size of the text feature vector \textcolor{black}{like} \citep{kochhar2014automatic}. As for the weighting function, we used the standard Term Frequency-Inverse Document Frequency (TF-IDF) \citep{manning2008schu} as it is commonly used in the literature \citep{gharbi2019classification,ouni2016multi,lin2013empirical,le2015rclinker}. 
The value for each N-Gram is proportional to its TF score multiplied by its IDF score. Thus, each preprocessed word in the commit message is assigned a value which is the weight of the word computed using this weighting scheme. TF-IDF gives greater weight (\textit{e.g.,} value) to words which occur frequently in fewer documents rather than words which occur frequently in many documents. 

\subsubsection{Model Training and Building}
In this phase, we performed the 10-fold cross-validation technique to assess the variability and reliability of the classifier. Specifically, for each of the classification methods, we combined the commit messages into a single large dataset. Then, we split the dataset into ten folds, where each fold contained an equal proportion of commit messages. Thereafter, we performed ten evaluation rounds with different testing dataset in which nine folds were used as training dataset and the remaining one of the ten folds is used as the testing dataset. We aggregated the results of the ten evaluation rounds and reported the average performance for each classifier.

\subsubsection{Classifier Selection and Model Evaluation}
Selecting the proper classifier for optimal classification of the commits is a rather challenging task \citep{fernandez2014we}. Best practices suggest that developers properly document their commits by providing a commit message along with every commit they make to the repository. These commit messages are typically written using natural language, and generally convey some descriptive information about the commit changes they represent. In this study, we are dealing with multiclass classification problem since the commit messages are categorized into six different types. Since we have a predefined set of categories (\ie refactoring types), our approach relies on supervised machine learning algorithms to assign each commit message to one category.  
Since it is very important to come up with an optimal classifier that can provide satisfactory results, several studies have compared various classifiers such as K-Nearest Neighbor (KNN), Naive Bayes Multinomial, Gradient Boosting Machine (GBM), and Random Forest (RF) in the context of commit classification into similar categories \citep{Levin:2017:BAC:3127005.3127016,levin2019towards, kochhar2014automatic}. These studies found that Random Forest (RF) often achieves high performance. We investigated each classifier in our study using common statistical measures (\textit{precision, recall, and F-measure}) of classification performance to compare each of them based on Azure Machine Learning (Azure ML) \citep{mund2015microsoft}. It is important to note that the calculation of F-measure for multiclass classification is not supported by Azure ML. Thus, we compute F-measure using the following formula: 

\begin{equation}
F = 2*\left(\frac{Precision * Recall}
       {Precision + Recall}\right)
\end{equation}
where Precision (P) and Recall (R) are calculated as follows:
\begin{equation}
P= \frac{tp}{tp+fp} , \;\;\;\;\;\;\; R= \frac{tp}{tp+fn}
\end{equation}

It is worth noting that a few models that we consider are inherently binary classifiers. In order to adjust for multiclass classification, each classifier applies the One-vs-All strategy for issues that require multiple output classes \citep{Lorena2009}. Thus, to ensure fairness, we use One-vs-All strategy for multiclass classification when using the following five classifiers: Gradient Boosted Machine (GMB), Support Vector Machine (SVM), Locally Deep SVM (LD-SVM), Averaged Perceptron Method (APM), and Bayes Point Machine (BPM). The remaining classifiers, consider in this study, are: Logistic Regression (LR), Random Forest (RF), Decision Jungle (DJ), and Neural Network (NN). Our experiment is conducted using Microsoft Azure Machine Learning platform (Azure ML)  \citep{mund2015microsoft}, as it provides a built-in web-service once the classification models are deployed.   \textcolor{black}{We provide, in Table \ref{Table:Parameters}, the default parameter values of the classification algorithms in our study replicability purposes.}

\begin{table}[h!]
\centering
\caption{Default Parameter Values for the Classification Algorithms.}
\label{Table:Parameters}
\begin{adjustbox}{width=1.0\textwidth,center}
\begin{tabular}{@{}llll@{}}
\toprule
\multicolumn{1}{l}{\textbf{Algorithm}} & \multicolumn{1}{l}{\textbf{Parameter}} & \multicolumn{1}{l}{\textbf{Description}} & \multicolumn{1}{l}{\textbf{Default Value}} \\ \midrule
\multirow{4}{*}{Random Forest} & n\_estimators          & Number of decision trees  & 8  \\ 
                               & max\_depth             & Maximum depth of the decision trees & 32   \\ 
                               & n\_samples\_leaf      & Number of random splits per node  & 128    \\ 
                               & min\_samples\_split    & Minimum number of samples per leaf node  & 1  \\
 \midrule
 \multirow{4}{*}{Logistic Regression} & optimiz\_tol &  Optimization tolerance &  1E-07 \\ 
          & 1\_weight                         &  L1 regularization weight &  1\\ 
                     & L2\_weight                       &  L2 regularization weight  &  1 \\ 
                     & memory\_L\_BFGS          &  Memory size for L-BFGS  &  20 \\
 \midrule
 \multirow{4}{*}{Gradient Boosted Machine} & max\_n\_leaf    & Maximum number of leaves per tree & 20    \\
         & min\_samples\_leaf        & Minimum number of samples per leaf node & 10 \\
                     & learning\_rate          & Learning rate & 0.2  \\
                     & n\_tree               & Number of trees constructed  & 100   \\
 \midrule
 \multirow{4}{*}{Decision Jungle} &  n\_estimators  & Number of decision directed acyclic graphs & 8  \\ 
          & max\_depth     & Maximum depth of the decision directed acyclic graphs  & 32\\ 
                     & max\_width     & Maximum  of the decision directed acyclic graphs & 128    \\ 
                     & n\_optimiz    & Number of optimization steps per decision directed acyclic graphs layer & 2048  \\ 
 \midrule
\multirow{2}{*}{Support Vector Classification} & n\_iter       & Number of iterations   & 1 \\        & Lambda    & Lambda & 0.001 \\ 
 \midrule
  \multirow{6}{*}{Locally Deep SVM}   &  max\_depth       &  Depth of the tree          &  3  \\ 
                    &  lam\_weight      &  Lambda weight &  0.1 \\ 
                    &  n\_theta    &  Lambda Theta  &  0.01   \\
                    &  n\_theta\_Prime       &  Lambda Theta Prime  &  0.01  \\
                  &  n\_sigmoid        &  Sigmoid sharpness  &  1  \\
                   &  n\_iter     &  Number of iterations &  15000   \\  
  \midrule
\multirow{5}{*}{Neural Network} & n\_nodes    & Number of hidden nodes  & 100  \\
                                 & learning\_rate & The learning rate & 0.1 \\
                          & n\_learning\_rate     & Number of learning iterations & 100   \\ 
                     & learning\_rate\_weights    & Initial learning weights diameter  & 0.1 \\ 
                     & momentum             & Momentum  & 0   \\  
\midrule
 \multirow{2}{*}{Average Perceptron Method}  & learning\_rate    & Learning rate  & 1   \\ 
          & m\_iter   & Maximum number of iterations & 10 \\
 \midrule
 \multirow{1}{*}{Bayes Point Machine}  & n\_training\_iter  & Number of training iterations &  30 \\   
 \bottomrule
\end{tabular}
\end{adjustbox}
\end{table}

\begin{table*}[h]

\centering
\caption{Performance of Each Model, in Terms of Precision (P), Recall (R), and F-measure (F1), per Refactoring Type (a set of 5,004 commits).} 
\label{Table:ClassifierScores_Details}
\begin{adjustbox}{width=1.0\textwidth,center}
\centering
\begin{tabular}{lrrrlrrrlrrr}
\hline
\multicolumn{4}{|c|}{\textit{\textbf{Random Forest}}} & \multicolumn{4}{c|}{\textit{\textbf{Logistic Regression}}} & \multicolumn{4}{c|}{\textit{\textbf{One-vs-All Gradient Boosted Machine}}} \\ \hline
\multicolumn{1}{c}{\textbf{Refactoring type}} & \multicolumn{1}{c}{\textbf{P}} & \multicolumn{1}{c}{\textbf{R}} & \multicolumn{1}{c|}{\textbf{F1}} & \multicolumn{1}{c}{\textbf{Refactoring type}} & \multicolumn{1}{c}{\textbf{P}} & \multicolumn{1}{c}{\textbf{R}} & \multicolumn{1}{c|}{\textbf{F1}} & \multicolumn{1}{c}{\textbf{Refactoring type}} & \multicolumn{1}{l}{\textbf{P}} & \multicolumn{1}{l}{\textbf{R}} & \multicolumn{1}{l}{\textbf{F1}} \\ \hline
Extract Method & 0.58 & 0.65 & \multicolumn{1}{r|}{0.62} & Extract Method & 0.63 & 0.64 & \multicolumn{1}{r|}{0.63} & Extract Method & 0.71 & 0.68 & 0.69 \\
Inline Method & 0.41 & 0.46 & \multicolumn{1}{r|}{0.44} & Inline Method & 0.43 & 0.48 & \multicolumn{1}{r|}{0.45} & Inline Method & 0.45 & 0.44 & 0.45 \\
Move Method & 0.57 & 0.67 & \multicolumn{1}{r|}{0.61} & Move Method & 0.57 & 0.61 & \multicolumn{1}{r|}{0.59} & Move Method & 0.61 & 0.66 & 0.63 \\
Pull Up Method & 0.41 & 0.31 & \multicolumn{1}{r|}{0.35} & Pull Up Method & 0.41 & 0.38 & \multicolumn{1}{r|}{0.40} & Pull Up Method & 0.42 & 0.41 & 0.42 \\
Push Down Method & 0.42 & 0.32 & \multicolumn{1}{r|}{0.36} & Push Down Method & 0.40 & 0.36 & \multicolumn{1}{r|}{0.38} & Push Down Method & 0.44 & 0.41 & 0.42 \\ 
Rename Method & 0.89 & 0.92 & \multicolumn{1}{r|}{0.91} & Rename Method & 0.93 & 0.87 & \multicolumn{1}{r|}{0.90} & Rename Method & 0.91 & 0.94  & 0.93 \\ \hline
\multicolumn{12}{l}{} \\ \hline
\multicolumn{4}{|c|}{\textit{\textbf{Decision Jungle}}} & \multicolumn{4}{c|}{\textit{\textbf{One-vs-All Support Vector Machine}}} & \multicolumn{4}{c|}{\textit{\textbf{One-vs-All Locally Deep SVM}}} \\ \hline
\multicolumn{1}{c}{\textbf{Refactoring type}} & \multicolumn{1}{c}{\textbf{P}} & \multicolumn{1}{c}{\textbf{R}} & \multicolumn{1}{c|}{\textbf{F1}} & \multicolumn{1}{c}{\textbf{Refactoring type}} & \multicolumn{1}{c}{\textbf{P}} & \multicolumn{1}{c}{\textbf{R}} & \multicolumn{1}{c|}{\textbf{F1}} & \multicolumn{1}{c}{\textbf{Refactoring type}} & \multicolumn{1}{c}{\textbf{P}} & \multicolumn{1}{c}{\textbf{R}} & \multicolumn{1}{c}{\textbf{F1}} \\ \hline
Extract Method & 0.54 & 0.66 & \multicolumn{1}{r|}{0.59} & Extract Method & 0.55 & 0.56 & \multicolumn{1}{r|}{0.55} & Extract Method & 0.54 & 0.54 & 0.54 \\
Inline Method & 0.40 & 0.43 & \multicolumn{1}{r|}{0.42} & Inline Method & 0.38  & 0.39 & \multicolumn{1}{r|}{0.39} & Inline Method & 0.35 & 0.35 & 0.35 \\
Move Method & 0.58 & 0.73 & \multicolumn{1}{r|}{0.65} & Move Method & 0.50 & 0.51 & \multicolumn{1}{r|}{0.50} & Move Method & 0.47 & 0.46 & 0.47  \\
Pull Up Method & 0.39 & 0.21 & \multicolumn{1}{r|}{0.27} & Pull Up Method & 0.37 & 0.36 & \multicolumn{1}{r|}{0.36} & Pull Up Method & 0.34 & 0.38 & 0.36 \\
Push Down Method & 0.38 & 0.27 & \multicolumn{1}{r|}{0.31} & Push Down Method & 0.37 & 0.38 & \multicolumn{1}{r|}{0.37} & Push Down Method & 0.41 & 0.39 & 0.40 \\ 
Rename Method & 0.90 & 0.96 & \multicolumn{1}{r|}{0.93} & Rename Method & 0.86 & 0.81 & \multicolumn{1}{r|}{0.84} & Rename Method & 0.85 & 0.78  & 0.81 \\  \hline
\multicolumn{12}{l}{} \\ \hline
\multicolumn{4}{|c|}{\textit{\textbf{Neural Network}}} & \multicolumn{4}{c|}{\textit{\textbf{One-vs-All Averaged Perceptron Method}}} & \multicolumn{4}{c|}{\textit{\textbf{One-vs-All Bayes Point Machine}}} \\ \hline
\multicolumn{1}{c}{\textbf{Refactoring type}} & \multicolumn{1}{c}{\textbf{P}} & \multicolumn{1}{c}{\textbf{R}} & \multicolumn{1}{c|}{\textbf{F1}} & \multicolumn{1}{c}{\textbf{Refactoring type}} & \multicolumn{1}{c}{\textbf{P}} & \multicolumn{1}{c}{\textbf{R}} & \multicolumn{1}{c|}{\textbf{F1}} & \multicolumn{1}{c}{\textbf{Refactoring type}} & \multicolumn{1}{c}{\textbf{P}} & \multicolumn{1}{c}{\textbf{R}} & \multicolumn{1}{c}{\textbf{F1}} \\ \hline
Extract Method & 0.58 & 0.50 & \multicolumn{1}{r|}{0.54} & Extract Method & 0.54 & 0.53 & \multicolumn{1}{r|}{0.53} & Extract Method & 0.49 & 0.46 & 0.48 \\
Inline Method & 0.37 & 0.37 & \multicolumn{1}{r|}{0.37} & Inline Method & 0.36  & 0.38 & \multicolumn{1}{r|}{0.37} & Inline Method & 0.33& 0.35 & 0.34 \\
Move Method & 0.50 & 0.44  & \multicolumn{1}{r|}{0.47} & Move Method & 0.45 & 0.48  & \multicolumn{1}{r|}{0.46} & Move Method & 0.40 & 0.49 & 0.44 \\
Pull Up Method & 0.36 & 0.35 & \multicolumn{1}{r|}{0.35} & Pull Up Method & 0.36 &  0.37 & \multicolumn{1}{r|}{0.36} & Pull Up Method & 0.36 & 0.35 & 0.36 \\
Push Down Method & 0.37 & 0.46 & \multicolumn{1}{r|}{0.41} & Push Down Method & 0.39 & 0.38 & \multicolumn{1}{r|}{0.39} & Push Down Method & 0.38 & 0.36  & 0.37 \\ 
Rename Method & 0.82 & 0.86 & \multicolumn{1}{r|}{0.84} & Rename Method & 0.85 & 0.81 & \multicolumn{1}{r|}{0.83} & Rename Method & 0.70 & 0.61 & 0.65 \\  \hline
\end{tabular}
\end{adjustbox}
\end{table*}

\section{Results \& Discussions}
\label{sec:results}

In this section, we assess the performance of our approach, and aim at answering the following research questions:

\begin{itemize}
\item \textbf{RQ1. (effectiveness)} \RQone
\item \textbf{RQ2. (baseline comparison)} \RQtwo
\item  \textcolor{black}{\textbf{RQ3. (terminology)} \RQthree}
\item  \textcolor{black}{\textbf{RQ4. (inconsistency)} \RQfour}

\end{itemize}


\textbf{Replication package.} We provide our comprehensive experiments package available in \citep{SAR2019WEB} to further replicate and extend our study.

\subsection{RQ1. \RQone}

Table \ref{Table:ClassifierScores_Details} reports the performance results of each classifier, in terms of precision, recall and F-measure, broken down per class, \ie refactoring type. 

According to Table \ref{Table:ClassifierScores_Details}, Random Forest (RF), Gradient Boosting Machine (GBM), and Logistic Regression (LR) are performing relatively higher than their competitor classifiers, in terms of F-measure, across the majority classes. We also observe that 
the GBM was able to achieve the highest average F-measure of 0.59, in comparison with RF and LR, whose F-measure is respectively 0.54 and 0.55. Random Forest and Boosting learning machines belong to the family of ensemble learning machines, and have typically yielded superior predictive performance mainly due to the fact that they both aggregate several learnings. As for Logistic Regression, the fact that Logistic Regression achieves comparable performance as Random Forest and Boosting can be explained by the fact that the underlying true model for the text data has an inherent structure that matches the logistic regression assumption. 

Overall, there is an interesting pattern that we can observe across all classifiers: there is an agreement between all models that the \textit{Rename Method} refactoring is the easiest to classify, with an F-measure starting from 0.65 (Bayes Point Machine) and reaching up to 0.93 (GBM). The \textit{Extract Method} refactoring classification was the second highest for all classifiers except Decision Jungle. Its F-measure varies from 0.48 (Bayes Point Machine) to 0.69 (GBM). Furthermore, we observe that for the \textit{Move Method} refactoring, the classifiers' performance varies between 0.46 (Averaged Perceptron Method) and 0.63 (GBM). As for the remaining classes, the performance of classifiers was similar and relatively low, when compared with the previous classes. For instance, the classifiers' performance, for the \textit{Inline Method} refactoring varies between 0.34 (Bayes Point Machine) and 0.45 (GBM). For the \textit{Pull-up Method} and the \textit{Push-down Method} refactorings, the highest F-measure scored across all classifiers was 0.42.
To gain a better understanding on why there exists such differences in the prediction between the refactoring types, we further analyzed the confusion matrix of the GBM classifier. During our qualitative analysis, we made the following observations:

\begin{table*}[h]
  \centering
	 \caption{Examples of Wrongly Predicted Commit Messages, by the Gradient Boosting Machine (GBM).}
	 \label{Table:example}
\begin{adjustbox}{width=1.0\textwidth,center}
\begin{tabular}{lllllll}\hline
\toprule
\bfseries Observation & \bfseries Ref. Operation & \bfseries Commit Message Example \\
\midrule
Similar Expression & Extract Method & 
\say{\textit{fcrepo-1029: \textbf{move} purge code \textbf{to} separate method}} \\ 
& Inline Method &  \say{\textit{ISQReader: \textbf{move} the dialog code \textbf{into} run() and tidy up}} \\ 
& Move Method & \say{\textit{\textbf{Move} send/receive code \textbf{from} SMTPSession \textbf{to} TextProtocolTester
[...]}} \\ 
& Pull Up Method & \say{\textit{HV-1239 \textbf{Moving} shared code \textbf{up} to CascadableConstraintMapping[...]}} \\ 
& Push Down Method & \say{\textit{\textbf{Move} group communication \textbf{down} to jvstm-ispn only [...]}} \\ 
\hline
Inadequate Expression & Extract Method & \say{\textit{\textbf{Merged} updateTopic and updateTopicInline.}} \\
& Inline Method & \say{\textit{\textbf{Extracting} transactions from HadoopArchiveFileSystem. [...]}}\\
& Move Method & \say{\textit{Improve code structure. \textbf{Added} tests.}}\\
& Pull Up Method & \say{\textit{\textbf{split} out into ERXAjaxContext so you can [...]}}\\
& Push Down Method & \say{\textit{\textbf{removed} deprecated method getConfigServer()}}\\
& Rename Method & \say{\textit{\textbf{Added} extended names for mixins.}}\\
\bottomrule
\end{tabular}
\end{adjustbox}
\vspace{-.3cm}
\end{table*}

\vspace{.2cm}
\noindent\textbf{Observation \# 1. Similar Expressions.} 

Our first observation relates to the terminology and keywords developers use to describe each refactoring type. We notice that \textit{Rename Method} has the highest accuracy across all classifiers because developers typically use the keyword \textit{rename} to describe renaming methods. However, for the other types, developers do not stick to how these types are named in the refactoring catalog, and use various terminologies, to describe them. We enumerate, in Table \ref{Table:example}, examples from messages belonging to \textit{Extract/Inline/Pull-up/Push-down Method} classes, and which were wrongly predicted as \textit{Move Method}. For instance, the process of extracting a method was described in one of the commits as "\textit{moving} purge code to a \textit{separate method}". While we can induce the extraction of the method, it was mislabeled by GBM classifier.

\vspace{.2cm}
\noindent\textbf{Observation \# 2. Inadequate Expressions.}

Occasionally, some messages contain keywords that are counter-intuitive to our model, resulting in a misclassification. Table \ref{Table:example} contains samples of misclassified commits, we report the correct label, while keywords that induced the wrong prediction are in bold. Let us take the following message: \say{\textit{Merged updateTopic and updateTopicInline}}, which documents inlining two methods, namely \texttt{updateTopic()} and \texttt{updateTopicInline()}, however, Refactoring Miner 
has detected an extraction of the method. To further understand this, we conducted a manual analysis of random samples. Our verification indicates that the keywords used by our model are not necessarily meant to document the underlying refactoring, as developers may document other changes performed in the commit. 


It is worth noting that a recent study has reported that developers do misuse refactoring-related terms in their documentations \citep{zhangpreliminary18}. Such cases will also hinder the accuracy of our prediction.



\begin{tcolorbox}
\textit{Summary.} The accuracy of refactoring prediction is not uniform across all types. Some types are easier to predict than others. The prediction results for \textit{Rename Method}, \textit{Extract Method}, and \textit{Move Method} were ranging from 63\% to 93\% in terms of F-measure. However, our model was not able to accurately distinguish between \textit{Inline Method}, \textit{Pull-up Method}, and \textit{Push-down Method}, as its F-measure was between 42\% and 45\%.
\end{tcolorbox}

\subsection{RQ2. \RQtwo}

We opt to test the keyword-based approach because it was used to identify refactoring commits in previous studies \citep{kim2014empirical,zhangpreliminary18,Ratzinger:2008:RRS:1370750.1370759,stroggylos2007refactoring,ratzinger2005improving,murphy2012we,Mauczka2012}. The keyword-based approach also measures the extent to which developers explicitly mention their refactoring operations in their commit messages. 

The keyword-based approach simply uses the following keywords, namely \say{\textit{extract}}, \say{\textit{inlin}}, \say{\textit{mov}}, \say{\textit{pull}}, \say{\textit{push}}, and \say{\textit{renam}}, to perform the prediction. Note that we manually check the results to remove any false matching, \eg for the keyword \textit{mov}, we filtered matchings like \textit{movie} and \textit{movement}.

Figures~\ref{Chart:Visualization of the Precision}, \ref{Chart:Visualization of the Recall}, and \ref{Chart:Visualization of the F1-measure for Different Classifiers} present the experimental results of our approach compared with the keyword-based prediction. Our approach provides an F-measure improvement across all refactoring types. One case in which the keyword-based approach could not detect the type of refactoring but the ML-based approach detects correctly is best illustrated in the following commit message: \say{\textit{Change name of `Decorator' to `Events'}}.  The keyword-based approach does not capture this message as it does not contain the keyword \say{\textit{renam}}. This is intuitive since the model has identified a set of keywords that were also used to indicate a given refactoring type. For example, if we refer to Table \ref{Table:Features}, the \textit{Inline Method} refactoring was found to be documented using various keywords such as \textit{combine}, \textit{gather}, and \textit{merge}. Similarly with the \textit{Extract Method} refactoring, whose documentation contained \textit{add}, \textit{create}, \textit{split}, and \textit{separate}.

It is worth noting that the highest performance of the keyword-based approach was achieved when predicting the \textit{Move Method} refactoring, being able to capture the vast majority of commits containing this type (true positives), along with many other commits containing mainly the \textit{Pull-up Method}, and \textit{Push-down Method} refactortings, because developers typically document them using the \say{\textit{move}} keyword, as we illustrated in Table \ref{Table:example}.

\begin{figure}[tbp] 
\centering
\begin{tikzpicture}
 \begin{scope}[scale=0.83]
  \centering
  \begin{axis}[
        ybar, axis on top,
        height=6cm, width=13cm,
        bar width=0.5cm,
        ymajorgrids, tick align=inside,
        major grid style={draw=white},
        enlarge y limits={value=.1,upper},
        ymin=0, ymax=100,
        axis x line*=bottom,
        axis y line*=right,
        y axis line style={opacity=0},
        tickwidth=0pt,
        enlarge x limits=true,
        enlarge x limits={abs=2cm}, 
        legend style={
            at={(0.5,-0.1)},
            anchor=north,
            legend columns=-1,
            /tikz/every even column/.append style={column sep=0.5cm}
        },
        ylabel={Precision (\%)},
        symbolic x coords={
           Extract, Inline, Move, Pull up, Push down, Rename},
       xtick=data,
       nodes near coords={
        \pgfmathprintnumber[precision=0]{\pgfplotspointmeta}
       }
    ]
    \addplot [draw=none, fill=blue!30] coordinates {
      (Extract,71)
      (Inline, 45)
      (Move, 61)
      (Pull up, 42)
      (Push down, 44)
      (Rename, 91)
      };
   \addplot [draw=none,fill=red!30] coordinates {
      
            (Extract,76)
      (Inline, 81)
      (Move,38)
      (Pull up, 69)
      (Push down,67)
      (Rename, 87)
};
    \legend{Our approach, Keyword-based}
  \end{axis}
  \end{scope}
  \end{tikzpicture}
  \caption{\textcolor{black}{Visualization of the Precision for Different Approaches.}}

\label{Chart:Visualization of the Precision}
\begin{tikzpicture}
 \begin{scope}[scale=0.83]
  \centering
  \begin{axis}[
        ybar, axis on top,
        height=6cm, width=13cm,
        bar width=0.5cm,
        ymajorgrids, tick align=inside,
        major grid style={draw=white},
        enlarge y limits={value=.1,upper},
        ymin=0, ymax=100,
        axis x line*=bottom,
        axis y line*=right,
        y axis line style={opacity=0},
        tickwidth=0pt,
        enlarge x limits=true,
        enlarge x limits={abs=2cm}, 
        legend style={
            at={(0.5,-0.1)},
            anchor=north,
            legend columns=-1,
            /tikz/every even column/.append style={column sep=0.5cm}
        },
        ylabel={Recall (\%)},
        symbolic x coords={
           Extract, Inline, Move, Pull up, Push down, Rename},
       xtick=data,
       nodes near coords={
        \pgfmathprintnumber[precision=0]{\pgfplotspointmeta}
       }
    ]
    \addplot [draw=none, fill=blue!30] coordinates {
      (Extract,68)
      (Inline, 44)
      (Move, 66)
      (Pull up, 41)
      (Push down, 41)
      (Rename, 94)
      };
   \addplot [draw=none,fill=red!30] coordinates {
      (Extract,30)
      (Inline,5)
      (Move,83)
      (Pull up, 6)
      (Push down,5)
      (Rename, 98)
     
      };

    \legend{Our approach, Keyword-based}
  \end{axis}
  \end{scope}
  \end{tikzpicture}
  \caption{\textcolor{black}{Visualization of the Recall for Different Approaches.}}
\label{Chart:Visualization of the Recall}
  
  \centering

\begin{tikzpicture}
  \centering
   \begin{scope}[scale=0.83]
  \begin{axis}[
        ybar, axis on top,
        height=6cm, width=13cm,
        bar width=0.5cm,
        ymajorgrids, tick align=inside,
        major grid style={draw=white},
        enlarge y limits={value=.1,upper},
        ymin=0, ymax=100,
        axis x line*=bottom,
        axis y line*=right,
        y axis line style={opacity=0},
        tickwidth=0pt,
        enlarge x limits=true,
        enlarge x limits={abs=2cm}, 
        legend style={
            at={(0.5,-0.1)},
            anchor=north,
            legend columns=-1,
            /tikz/every even column/.append style={column sep=0.5cm}
        },
        ylabel={F-measure (\%)},
        symbolic x coords={
            Extract, Inline, Move, Pull up, Push down, Rename},
       xtick=data,
       nodes near coords={
        \pgfmathprintnumber[precision=0]{\pgfplotspointmeta}
       }
    ]
    \addplot [draw=none, fill=blue!30] coordinates {
      (Extract, 69)
      (Inline, 45)
      (Move, 63)
       (Pull up, 42)
      (Push down, 42)
      (Rename, 93)
      };
   \addplot [draw=none,fill=red!30] coordinates {
      (Extract,44)
      (Inline, 9)
      (Move,53)
      (Pull up,12)
      (Push down, 10)
      (Rename, 92)
      };
 
    \legend{Our approach, Keyword-based}
  \end{axis}
  \end{scope}
  \end{tikzpicture}
  \caption{\textcolor{black}{Visualization of the F-measure for Different Approaches.}}
\label{Chart:Visualization of the F1-measure for Different Classifiers}
 \end{figure}

\begin{table}[h]
\begin{center}
\caption{Relevant Features per Class.}
\label{Table:Features}
\begin{adjustbox}{width=1.0\columnwidth,center}
\begin{tabular}{llllll}\hline
\toprule
\bfseries Extract & \bfseries Inline & \bfseries Move  & \bfseries Pull Up & \bfseries Push Down & \bfseries Rename  \\
\midrule
 Add &  Combine   &  Move &  Move &  Move &  Change  \\
 Create     &  Gather  & Add  &  Pull &  Push &  Fix  \\
 Extract &  Inline  &  &  Shift & Reduce  &  Improve  \\
 Move &  Merge  &    &       &  Remove     &  Rename  \\
 Separate &  Move &     &     &    &  Update \\
 Split   &         &     &     &  &   \\
 Break up & \\
\bottomrule
\end{tabular}
\end{adjustbox}
\end{center}
\end{table}

\begin{tcolorbox}
\textit{Summary.} The keyword-based approach performs significantly lower than ML models. It assumes that developers are familiar with the catalog of refactorings, or refactoring types being offered in the IDEs. Our findings show that developers tend to document refactoring using the same set of patterns.
The keyword-based approach scored relatively better performance for the \textit{Rename Method} type because its keyword (\ie rename) is intuitive, in contrast with other types, such as \textit{Inline Method} and \textit{Push-down Method}.
\end{tcolorbox}

\subsection{RQ3. \textcolor{black}{\RQthree}}

\textcolor{black}{This research question examines the textual content of the commit messages to determine the frequent refactoring types-related terminology developers utilize when documenting their refactoring activity. In this RQ, we utilize natural language processing techniques, more specifically bigram analysis, to extract the frequent bigrams developers utilize in describing their refactoring activity for each refactoring type considered in our study. Bigrams are a sequence of two adjacent words in a sentence; in this instance, the commit messages. We also look at trigrams to locate sets of common terms. Unlike unigrams, bigrams and trigrams provide a certain level of context for terms, which helps our analysis by reducing the chance of making false presumptions. Before our extraction, we first run Refactoring Miner in order to identify commits containing refactorings from each type of refactoring operations considered in this study as discussed in Section \ref{sec:methodology}.}

\textcolor{black}{Upon a closer inspection of the refactoring patterns in Tables \ref{Table:doc1}, \ref{Table:doc2}, and \ref{Table:doc3}, we have made several observations: (1) the keywords and phrases used in renaming refactorings are the most discriminative, indicating that these terms are strongly associated with the action of renaming, (2) the patterns used for extract refactorings are associated with the motivation behind refactoring, \eg remove duplication, improve clarity, and improve reusability, (3) for move, pull up, and push down, developers used the term “move” interchangeably as the main action of these refactoring operations involve moving the code elements, and (4) the terms used in inlining refactorings are limited as developers mainly used specific keywords to demonstrate the action.}

\begin{table}[htbp]
\begin{center}
\caption{\textcolor{black}{Relevant Terms per Refactoring Types.}}
\label{Table:doc1}
\begin{adjustbox}{width=1.0\columnwidth,center}
\begin{tabular}{llllll}\hline
\toprule
\bfseries Rename & \bfseries Extract \\
\midrule
alter* method name for more consistency & add* a new method \\
better method name & add* method\\
chang* method name & add* new [] function\\
chang* method name for clarity & add* new method\\
chang* method name for consistency & add* several methods\\
chang* some method name & add* some convenience functions\\
chang* test method name & add* the [] method\\
chang* the method name & add* the method []\\
chang* the name & break* up the jumbo methods\\
clarif* method name & brok* up long methods into a bunch of smaller methods\\
clean* up method name & brok* up the [] method into a separate []\\
correct* a method name & creat* a higher level [] method\\
correct* method name & creat* a new method\\
fix* a typo in a method name & creat* method\\
fix* confusing method name & creat* separate method\\
fix* inconsistent method name & extract* common code from \\
fix* incorrect method name & extract* a few methods out\\
fix* method name & extract* a method\\
fix* method name conflict & extract* abstract method\\
fix* method name typo & extract* common code\\
fix* misspelled method name & extract* common method\\
fix* several method names & extract* method\\
fix* spelling for method name & extract* out a method\\
fix* typo in method name & extract* out function\\
improv* method name & extract* out the method\\
improv* the name & extract* some methods\\
made the method name a bit more explicit & extract* some methods for code clarity sake\\
method name chang* &  extract* the [] method from []\\
method name fix* & extract* some stuff to a method\\
method name improv* &fix* for method code size\\
method name refactor* & mov* [] into separate methods\\
method names in tests changed & refactor* duplicate code into separate method\\
minor change to method name & refactor* some methods\\
minor refactorings to method name & refactor*: Introduc* a method\\
modif* test method name & separat* [] from []\\
more meaningful method name &  separat* a method\\
normaliz* getter method name & split* [] into separate methods\\
polish test method name & split* into separate functions\\
refactor* method name & split* into smaller pieces first\\
refactor* some method names & split* into some smaller assert to reuse\\
renam* factory methods & split* the [] into component parts for clarity\\
renam* for clarification & split* the [] method in several sub-methods\\
renam* for clarity & split* the [] method into a []\\
renam* for consistency & split* the code into [] and []\\
renam* method &  split* the HUGE generate method into different methods\\
renam* method name & split* up\\
renam* misleading method name & split* up [] a bit more neatly\\
renam* of code & split* up a complex method\\
renam* of component & split* up the [] method\\
renam* of function name & split* up the [] method into some methods\\
renam* some internal variables and methods &\\
renam* some methods &\\
renam* the method &\\
shorten* method name &\\
simplif* user method name &\\
solv* typo in method name &\\
standardization of method name &\\
tid* up method naming &\\
tid* up test method name &\\
unif* execution method name &\\
uniformiz* method name &\\
updat* method name &\\
updat* the test name &\\
using more correct method name & \\

\bottomrule
\end{tabular}
\end{adjustbox}
\end{center}
\end{table}

\begin{table}[htbp]
\begin{center}
\caption{\textcolor{black}{Relevant Terms per Refactoring Types (cont.).}}
\label{Table:doc2}
\begin{adjustbox}{width=1.0\columnwidth,center}
\begin{tabular}{llllll}\hline
\toprule
\bfseries Move & \bfseries Inline \\
\midrule
mov* [] to [] & add* methods for merge operation \\
mov* [] to new method &  combin* method\\
mov* all code into the only implementing class & consolidat* methods \\
mov* all utility methods into the same class & consolidat* some code\\ 
mov* around some methods & delet* unused method\\
mov* code around & inlin* helper methods\\
mov* formerly static methods to new & inlin* method\\
mov* from [] to [] & inlin* method only called once \\
mov* into & inlin* private method\\
mov* method & inlin* some methods\\
mov* out of & inlin* some trivial method\\
mov* some & inlin* the simplest method\\
mov* some code into a static utility method & merg* [] and [] into 1 method \\
mov* some methods and/or classes around & merg* [] and [] methods\\
mov* some methods to & merg* code into static method\\
mov* some of it's responsibilities out to other classes & merg* refactoring\\
mov* some of the methods into a class & merg* some code simplification\\
mov* some static methods to Utils &  more cleanup and merge resolution\\
mov* some stuff & refactor* [] into []\\
mov* static methods to a util class & refactor*: remov* some unused methods\\
mov* stuff out of the & remov* unused methods\\
mov* the [] & simplif* things my inlining both the method and the argument \\
mov* the implementation of the methods to & some consolidation of methods\\
mov* the method tests in their own class & useless method inlined\\
mov* the methods \\
mov* the notion of [] from [] to [] &\\
mov* to [] & \\
mov* util methods & \\
refactor* : mov* code &\\
refactor* out the methods into separate class &\\
refactor* some methods &\\
refactor* some methods to external helper class &\\
refactor* the code to move the [] to the [] &\\
refactor* to move the [] to [] &\\
refactor*: Move helper method to helper class &\\
refactor*: move to a helper method &\\
some static methods were moved from [] to [] & \\
\bottomrule
\end{tabular}
\end{adjustbox}
\end{center}
\end{table}

\begin{table}[h]
\begin{center}
\caption{\textcolor{black}{Relevant Terms per Refactoring Types (cont.).}}
\label{Table:doc3}
\begin{adjustbox}{width=1.0\columnwidth,center}
\begin{tabular}{llllll}\hline
\toprule
\bfseries Pull Up & \bfseries Push Down \\
\midrule
bunch of methods pulled up & chang* to shift functions\\
mov* common code in & minimal code duplication\\
mov* common code into & mov* common parts of\\
mov* common code to & mov* references to [] and [] into subclass\\
mov* more methods to & mov* test sections out of\\
mov* the common unit test setup to a base class & mov* [] from superclass\\
mov* the implementation to the superclass & mov* [] implementations into subclasses\\
mov* to & mov* some methods off [] onto a [] subclass\\
pull* to class level & push [] into []\\
pull* common & push to method level\\
pull* from & push* down\\
pull* from a specified & push* down to\\
pull* out & push* entities around\\
pull* out some common functionality & push* the [] code down into the \\
pull* out test methods into common area & push* to\\
pull* reusable & push* some stuff down from \\
pull* reusable code out of & reduc* the amount of implementation-specific code\\
pull* to & remov* dependency on\\
pull* up & remov* duplicate\\
pull* up common methods & remov* redundant\\
pull* up more properties to the base type & remov* redundant functions\\
pull* up some functionality from & stuff moved to separate\\
pull* up some methods &\\
pull* up to &\\
pull* out common code &\\
refactor* to "pull up" &\\
shift* further method to parent & \\
\bottomrule
\end{tabular}
\end{adjustbox}
\end{center}
\end{table}

\begin{tcolorbox}
\textit{Summary.} \textcolor{black}{Developers discriminate against different refactoring types through human language descriptions. The terminology used in rename refactorings are the most discriminative, indicating that these terms are strongly associated with the action of renaming.}
\end{tcolorbox}

\subsection{RQ4. \textcolor{black}{\RQfour}}

\textcolor{black}{Although our approach attempted to thoroughly predict method-level refactoring types, several inconsistency types between source code and documentation might occur. Several studies \citep{arnaoudova2016linguistic,fakhoury2019measuring,kim2016automatic} have identified and detected recurring poor practices related to inconsistencies among the documentation and implementation of the code elements. Because such inconsistencies can affect software comprehensibility and maintainability, this research question aims at exploring the frequency of different inconsistency types that might help in reporting any early inconsistency between refactoring types detected by refactoring detector tools and their documentation. Specifically, we are studying the following inconsistency types:}

\vspace{.2cm}
\noindent\textbf{\textcolor{black}{Case \# 1. Refactoring of type A is detected based on the source code but the description does not correspond to any refactoring.}}

\textcolor{black}{To obtain the data for this type of inconsistency, we need to add a set of commits in which the documentation does not correspond to any type of refactorings considered in this study. We started by randomly selecting 834 refactoring commits detected by Refactoring Miner while making sure no specific documentation about refactoring is reported. For example, we excluded the terms \say{\textit{extract}}, \say{\textit{inlin}}, and \say{\textit{mov}} since these terms correspond to the method-level refactoring operations. The 834 commits equated to the number of commits per refactoring type, as shown in Table \ref{Table:Instances per class (train, test)}. We then had to manually examine the list of commits to determine their appropriateness for this analysis. Next, we built a new model by considering adding this set to the training data with a \say{None} label. Since RQ1 shows that the GBM was able to achieve the highest average F-measure of 0.59, we used the GBM for our model, and we achieved the average F-measure of 0.58. Using a confidence level of 99\% and
 an interval of 5\%, we constructed a sample size of 588 commits for the manual analysis. The majority of these commits (85.03 \%) indicated there is a consistency between the refactoring detector and the model prediction, whereas a minority of these commits (14.96\%) shows inconsistent results.}
 
\textcolor{black}{The main challenge that we observed across various commits, is the tendency of developers to provide a high-level description of their refactoring, through the use of general expressions and patterns, such as \textit{refactor}, \textit{restructure}, and \textit{code clean up}, etc. Such patterns cannot be framed into one single type, \ie they can be used to describe all refactoring types. The following example demonstrates such a case:}

\begin{center}
\fbox{\parbox{\dimexpr\linewidth-2\fboxsep-2\fboxrule\relax}{\centering
``Just cleaned up the code a bit.''
}}

\captionof{Quote}{\textcolor{black}{Inconsistency type (Case \# 1)} \label{Quote:case1}}
\end{center}

\textcolor{black}{This phenomenon of using high level description to document low-level changes is also observed frequently in bug fix commit messages, where text messages would just contain the popular pattern of "\textit{fix bug X}". However, this is less problematic in the context of bugs because developers can still use the bug number (\textit{}X) to locate the corresponding bug report, and so access the bug’s proper documentation in the bug report. Whereas, for refactoring documentation, this is a persistent problem since without providing the rationale and the appropriate explanation of the change, there is no way to trace back such information anywhere in the project.}

\textcolor{black}{In practice, developers perform refactorings as singular transformations and in conjunction with other refactorings (\ie batch or composite refactorings). Previous studies (\eg \citep{bibiano2020does}) explored how single or composite refactorings contribute to the code smell removal or internal quality attribute improvement. Since developers perform these kinds of refactorings at the source code level, we expect that developers apply such practice of single or multiple transformation types at the documentation level on real development practices. Our previous studies on refactoring documentation showed that developers self-affirmed the action of refactoring in both open source (\eg \citep{alomar2019can,alomar2020toward,alomar2021we}) and industry \citep{alomar2021icse} at different levels of granularity including the high-level and fine-grained descriptions. A previous study \citep{yamashita2020changebeadsthreader} on tailoring untangled changes pointed out that developers often mix changes in different intentional tasks in one comment. The authors proposed an approach that regards a sequence of fine-grained changes that are about to be committed as a single commit by developers to merge and split change clusters to support the manual tailoring of untangling changes.} 

\textcolor{black}{From a practical point of view, researchers and practitioners can benefit from the proposed model to detect inconsistency types between refactoring detectors at the source code and documentation level, and to accelerate code review process since recent studies expressed the need to improve the quality of documentation for refactoring and non-refactoring changes \citep{alomar2021icse,ebert2021exploratory}}.

\vspace{.2cm}
\noindent\textbf{\textcolor{black}{Case \# 2. Refactoring of type A is detected based on the description but the source code change does not correspond to any refactoring.}}

\textcolor{black}{To perform our analysis, we need to include a set of commits that do not correspond to any refactoring operations and then feed this set into the training data with a \say{None} label. Thus, after running Refactoring Miner on a set of commits, we randomly selected 834 non-refactoring commits as indicated by Refactoring Miner. The selection of 834 commits was due to the count of refactoring types (see Table \ref{Table:Instances per class (train, test)}). We then built a model considering adding the set of non-refactoring commits in the training data. \textcolor{black}{Similarly to Case} \#1, we consider using the GBM for the newly created model, and we achieve the average F-measure of 0.56. To better understand the nature of this type of inconsistency, we performed a manual validation of 588 commits from the test data, this sample corresponds to a confidence level of 99\% and a confidence interval of 5\%. The majority of the commits (436 instances or 74.14\%) shows an agreement between the results obtained from the tool and our model, and (152 instances or 25.85\%) illustrates the disagreement case.}

\textcolor{black}{\citep{soares2020relation} reported that such type of inconsistency might indicate that developers apply refactorings that are different from refactorings defined by  \citep{Fowler:1999:RID:311424}.} \textcolor{black}{Moreover, we observe in our study that developers are documenting what they consider to be refactoring in non source code files. These files include configuration files, maven file, or database are not associated with refactoring operations detected by the tool even though the description contains refactoring operation-related keywords. The following example demonstrates such a case:}

\begin{center}
\fbox{\parbox{\dimexpr\linewidth-2\fboxsep-2\fboxrule\relax}{\centering
``Renamed table. The same table name was used in another test, which made this test fail when running all tests.''
}}
\captionof{Quote}{\textcolor{black}{Inconsistency type (Case \# 2)} \label{Quote:case2}}
\end{center}

\textcolor{black}{Such changes would not be detected by Refactoring Miner or any other detection tool because these tools are conceived to operate on only source files. Interestingly, our model results show that developers would also perform what they call refactoring on other files. If we refer to the original definition of refactoring, these changes may not be necessarily considered as refactorings, but with the rise of continuous integration, and infrastructure as service, many non-source files are now evolving as part of the project's ecosystem. These files undergo maintenance and evolution as well (updating dependencies, changing configurations, etc.). Therefore, there is a need for the refactoring community to properly taxonomize changes to these files, and evolve its toolset to detect them as well.} \textcolor{black}{Existing studies on configuration files have focused on the interactions between Java and XML configuration files \citep{chen2008toward}, the identification and detection of CI configuration bad practices that violate the best practices in CI configuration files (\eg redirecting scripts into interpreters, bypassing security checks, and using commands in an incorrect phase) and the prevalence of these anti-patterns in CI specifications \citep{zampetti2020empirical,gallaba2018use}. Since refactoring on other files is under research, future CI research and tooling needs to focus on the development of automated CI anti-pattern detectors and refactoring recommenders, and avoid the consequences of misusing CI features.}


\vspace{.2cm}
\noindent\textbf{\textcolor{black}{Case \# 3. Refactoring of type A is detected based on the source code, refactoring of type B is detected based on the description and A is different from B.}}

\textcolor{black}{Previous studies investigated the case when there is a disagreement between source code and its documentation in the context of programming misconception \citep{swidan2018programming}, linguistic anti-patterns \citep{arnaoudova2016linguistic}, bug localization \citep{fakhoury2019measuring}, and code review \citep{ebert2021exploratory}. In their study on misconceptions in programming education for school students, \citep{swidan2018programming} observed that younger learners hold common programming misconceptions that cause them to make errors. The authors recommended developing intervention methods to catch those misconceptions as early as possible. Further, \citep{arnaoudova2016linguistic} investigated developers' perception of linguistic anti-patterns and developed a catalog of 17 types of linguistic anti-patterns related to inconsistencies, findings that the majority of the participants perceive linguistic anti-patterns as
poor practices and must be avoided. \citep{fakhoury2019measuring} showed that inconsistencies in the source code have a significant effect on cognitive load, success, and time spent on program comprehension. More recently, \citep{ebert2021exploratory} discussed how developers deal with confusion in code reviews caused by unclear commit messages and lack of documentation. According to their survey with developers, one of the most frequent reasons for confusion is lack of documentation and missing code change rationale.}

\textcolor{black}{Since the presence of inconsistencies can mislead developers, we aim to investigate this phenomenon.} \textcolor{black}{For this type of inconsistency, we randomly selected 588 refactoring commits to check the percentage of the agreement and the mismatch between refactoring types detected by the Refactoring Miner and our model. This quantity roughly equates to a sample size with a confidence level of 99\% and a confidence interval of 5\%. We then run our deployed model on these commits in order to compare our results with that obtained by the Refactoring Miner. The result shows that the inconsistency case represents 60.20\% of the commits whereas only 39.79\%  of the commits are consistent.}

\textcolor{black}{Concerning our manual analysis, we observe that developers provide inadequate description of the code changes. The following example demonstrates such a case in which the tool detected composite refactoring operations \ie \textit{Extract}, \textit{Rename}, and \textit{Move} whereas our model predicted the commit as \textit{Extract} based on the description:}

\begin{center}
\fbox{\parbox{\dimexpr\linewidth-2\fboxsep-2\fboxrule\relax}{\centering
``Extract BindingHelper for re-use in wizards.''
}}
\captionof{Quote}{\textcolor{black}{Inconsistency type (Case \# 3)} \label{Quote:case3}}
\end{center}

\textcolor{black}{Our analysis for the three types of inconsistency shows that there is a need to improve the quality of refactoring documentation, and encourage the invention of the refactoring documentation generator. This offers a valuable opportunity to improve and standardize the format of the documentation. We believe that by combining the documentation with the state-of-the-art refactoring detectors, we can better understand the applied refactoring. For future work, we plan to perform an in depth study and extensive manual low-level source code inspection to better understand the phenomenon (\ie inconsistency cases).} 

\begin{tcolorbox}
\textit{Summary.} \textcolor{black}{Our model can work in conjunction with refactoring detectors \citep{tsantalis2018accurate,silva2017refdiff} in order to report any early inconsistency between refactoring types and their documentation.}
\end{tcolorbox}

\section{Research Implications}
\label{sec:Implication}
The main implications of this study are as follows:
\begin{enumerate}
    \item While existing studies, in classifying code changes using their commit messages \citep{Levin:2017:BAC:3127005.3127016,gharbi2019classification,levin2019towards}, have been achieving relatively higher accuracies in comparison with our model, this reveals a lack of refactoring documentation \textit{culture}, unlike documenting other code changes such as, API migration, bug fixes, and feature updates. However, the end goal our model is not to detect refactorings, but to work in conjunction with refactoring detectors \citep{tsantalis2018accurate,Silva:2016:WWR:2950290.2950305} in order to report any early inconsistency between refactoring types and their documentation. This is useful not only to improve the quality of documentation, which has been found to be lacking when it comes to describing code changes \citep{treude2020beyond}, but also to improve the understandability of code changes for code review and evolution purposes. For instance, a recent study has found that revealing more details about refactoring, such as types and intents, helps in facilitating its acceptance in code reviews \citep{bibiano2020does}. 
    
    \item  The words and phrases used in rename refactorings are the most discriminative, indicating that these terms are strongly associated with the action of renaming. Future work to help document rename refactorings, which are shown to be under-documented at between 1 and 6\% of the time \citep{arnaoudova:2014, peruma2020contextualizing}, can use our approach to determine what keywords they should use, or recommend to developers, when generating commit messages.
    
    \item Refactorings are generally associated with a specific set of keywords and phrases found in commit messages. However, there is also a significant amount of ambiguity in the way words are used; particularly for pull-up and push-down refactorings. A system which recommends how to document refactorings can reduce this confusion and the keywords that we discuss in this work are a strong starting point for determining what phrases should be used to reduce ambiguity.
    
    \item Our approach can be used to study the discriminative terms found in commit messages and can be used to detect the common words and phrases which describe different types of refactorings. In this study, we used this approach on a large number of systems but it could also be used on singular systems to detect project-specific ways of describing refactorings; further bolstering any future recommendation system's ability to tailor recommended commit messages/keywords to a specific project.
    
    \item Our study helps us understand refactoring documentation practices that trigger the need to explore the motivation behind refactoring. The study helps future developers to follow best documentation practices and improve the quality of the refactoring documentation. Further, the refactoring motivations tell the opinion of developers, so it is important for managers to learn developers' opinions and feelings especially for distributed software development practices. If developers do not document, managers will not know their intention. Since software engineering is a human-centric process, it is important for managers to understand the people's intention to work on the team through their documentation.
\end{enumerate}

\section{Threats to Validity}
\label{sec:threats}

In this section, we describe potential threats to validity of our research method, and the actions we took to mitigate them.

\textbf{Internal Validity.} \textcolor{black}{Our analysis is mainly threatened by the accuracy of the Refactoring Miner tool because the tool may miss the detection of some refactorings. However, previous
studies \citep{tsantalis2018accurate,Silva:2016:WWR:2950290.2950305} report that Refactoring Miner has high
precision and recall scores (\textit{i.e.}, a precision of
98\% and a recall of 87\%) compared to other state-of-the-art refactoring detection tools and is frequently
utilized in refactoring studies (\textit{e.g.,}\citep{aniche2020effectiveness,chavez2017does,peruma2020contextualizing,alomar2019can,alomar2020toward,alomar2020relationship,alomar2020developers,alomar2021refactoringreuse}).  A recent survey \citep{tan2019survey} compares several refactoring detection tools and shows that Refactoring Miner is currently the most accurate refactoring detection tool, which gives us confidence in using the tool.}

\noindent\textbf{Construct Validity.} 
Since our approach heavily depends on commit messages, we used well-commented Java projects when performing our study. Thus, the quality and the quantity of commit messages might have an impact on our findings. Another important limitation concerns the size of the dataset used for training and evaluation. The size of the used dataset was determined similarly to previous commit classification studies, but we are not certain that this number is optimal for our problem. It is better to use a systematic technique for choosing the size of the evaluation set. Another threat to validity can be related to the list of keywords that we used to identify set of commits for keyword-based approach as developers might use other keywords when documenting refactoring. However, the impact of this threat was limited to the refactoring operation-related keywords detected by Refactoring Miner. 

\noindent\textbf{External Validity.} 
The first threat relates to the commits that are extracted
only from open source Java projects. Our results may not generalize to commercially developed projects, or to other projects using different programming languages. Further, since a commit message could potentially belong to multiple refactoring types, our model does not consider such cases. However, exploring how to
automatically classify commits into this kind of hybrid categories is an interesting direction for future work.

\vspace{-.2cm}

\section{Conclusion}
\label{sec:conclusion}

In this paper, we formulated the prediction of refactorings as a multiclass classification problem, \textit{i.e.,} classifying refactoring commits into six method-level refactoring operations, applying nine supervised machine learning algorithms. We compared the performance of our approach to the keyword-based baseline and our results show that our approach outperforms the keyword-based approach. \textcolor{black}{Specifically, our main findings show that (1) the prediction results for \textit{Rename Method}, \textit{Extract Method}, and \textit{Move Method} were ranging from 63\% to 93\% in terms of F-measure. However, our model was not able to accurately distinguish between \textit{Inline Method}, \textit{Pull-up Method}, and \textit{Push-down Method}, as its F-measure was between 42\% and 45\%, (2) the keyword-based approach performs significantly lower than ML models, (3) developers discriminate against different refactoring operations through human language descriptions, and (4) there is a need to improve the quality of refactoring documentation and encourage the invention of the refactoring documentation generator.}

In the future, we plan to study the applicability of our approach to other
projects developed in different programming languages, and to other domains, \ie consider using commit messages written in different programming languages to predict refactoring and compare findings. We also plan to use the extension of Refactoring Miner \citep{tsantalis2020refactoringminer} that supports low-level refactorings. Another interesting research direction is to investigate if our approach can be applied to statement-level refactoring (\eg \textit{Extract Variable}). \textcolor{black}{Additionally, since a commit message could potentially belong to multiple categories (\textit{e.g.}, \textit{Extract Method} and \textit{Move Method}), future research could usefully apply multi-label classification to automatically classify commits into this kind of hybrid categories}. Further, although we used commit messages as our primary source of text, our approach is not restricted to a specific source of textual information. In our future work, we can test our approach using other types of information, including
issue descriptions.


\section{Acknowledgments}

This material is based on work supported by the National Science Foundation under Grant No. 1757680.

\bibliographystyle{model5-names}
{\footnotesize\bibliography{references.bib}}

\end{document}